\documentclass[12pt]{article}
\pdfoutput=1	
\usepackage[utf8]{inputenc}
\usepackage{array} 
\usepackage{amsmath,amssymb}
\usepackage{graphicx}
\usepackage[pdftex]{hyperref}
\usepackage{caption}
\DeclareGraphicsExtensions{.eps,.bmp,.wmf,.jpg}
\numberwithin{equation}{section}
\def\be{\begin{equation}}
\def\ee{\end{equation}}

\def\bea{\begin{eqnarray}}
\def\eea{\end{eqnarray}}

\title{Slow-Roll Inflation in Scalar-Tensor Models}
\author{L.N. Granda\thanks{luis.granda@correounivalle.edu.co} ,\, D. F. Jimenez\thanks{jimenez.diego@correounivalle.edu.co}\\{\it Departamento de Fisica, Universidad del Valle}\\{\it A.A. 25360, Cali, Colombia}}
\date{}
\begin{document}
\maketitle


\begin{abstract}
The linear and quadratic perturbations for a scalar-tensor model with non-minimal coupling to curvature, coupling to the Gauss-Bonnet invariant and non-minimal kinetic coupling to the Einstein tensor are developed. The quadratic action for the scalar and tensor perturbations is constructed and the power spectra for the primordial scalar and tensor fluctuations are given. A consistency relation that is useful to discriminate the model from the standard inflation with canonical scalar field was found. For some power-law potentials it is shown that the Introduction of additional interactions, given by non-minimal, kinetic  and Gauss-Bonnet couplings, can lower the tensor-to-scalar ratio to values that are consistent with latest observational constraints, and the problem of large fields in chaotic inflation can be avoided. 
\end{abstract}

\maketitle


\section{\label{intro}Introduction}
The improvement in the quality of the cosmological observations of the last years \cite{planck13, planck15, planck18, bkp} has reinforced the theory of cosmic inflation \cite{guth, linde, steinhardt}. The inflationary theory gives by now the most likely scenario for the early universe,  since it provides the explanation to flatness, horizon and monopole problems, among others, for the standard hot Bing Bang cosmology \cite{revlinde,liddle, riotto, lyth, mukhanov, baumann, nojirioo}. In other words, the inflation can set the initial conditions for the subsequent hot Big Bang, by eliminating the fine-tuning condition needed for solving the horizon, flatness and other problems. Besides that, the quantum fluctuations during inflation could provide the seeds for the large scale structure and the observed CMB anisotropies \cite{starobinsky1, mukhanov1, starobinsky3, hawking, starobinsky2, guth1, bardeen0, bardeen}. In particular, inflation allows as to understand how the scale-invariant power spectrum can be generated, though it does not predict an exact scale invariant but nearly scale invariant power spectrum \cite{bardeen}. The deviation from scale invariance is connected with the microphysics description of the inflationary theory which is still incomplete.\\
The simplest and most studied model of inflation consists of minimally-coupled scalar field with flat enough potential to provide the necessary conditions for slow-roll \cite{linde, steinhardt}. But the inflation scenario can be realized in many other models like non-minimally coupled scalar field \cite{futumase, unruh, barrow1, barrow2, bezrukov}, kinetic inflation \cite{picon}, vector inflation \cite{ford, koivisto, mukhanov2}, inflaton potential in supergravity \cite{kawasaki, davies, kallosh}, string theory inspired inflation \cite{soda1, soda2, maldacena, kallosh1, soda3, baumann2}, 
Dirac-Born-Infeld inflation model \cite{silver, silver1, chen, easson}, $\alpha$-attractor models originated in supergravity \cite{linde3, ferrara, sergei1, dimopoulos, akrami}. Apart from the DBI models of inflation, another class of ghost-free models has been recently considered, named "Galileon" models \cite{nicolis, deffayet}. In spite of the higher derivative nature of these models, the gravitational and scalar field equations contain derivatives no higher than two. The effect of these Galileon terms is mostly reflected in the modification of the kinetic term compared to the standard canonical scalar field, which in turn can improve (or relax) the physical constraints on the potential. In the case of the Higgs potential, for instance, one of the effects of the higher derivative terms is the reduction of the self coupling of the Higgs boson, so that the spectra of primordial density perturbations are consistent with the present observational data \cite{kamada, ohashi} (which is not possible within the standard canonical scalar field inflation with Higgs potential). Different aspects of Galilean-inflation have been considered in \cite{kamada, ohashi, yokoyama, mizuno, burrage, kobayashi1}. A particular and important case belonging to the above class of models is the scalar field with kinetic coupling to the Einstein tensor \cite{capozziello, sushkov, granda1, granda2} whose application in the context of inflationary cosmology has been analyzed in \cite{capozziello1, germani, granda3, stsujikawa, nanyang, gansu}.\\ 
\noindent In the present paper we consider a scalar-tensor model with non-minimal coupling to scalar curvature, non-minimal kinetic coupling to the Einstein tensor and coupling of the scalar field to the Gauss-Bonnet 4-dimensional invariant, to study the slow-roll inflation and the observable magnitudes, the scalar espectral index and the tensor-to-scalar ratio, derived from it. These interaction terms have direct correspondence with terms presented in Galileon theories \cite{kobayashi1, edcopeland}. 
This model is the simplest and more general scalar-tensor theory (whose Lagrangian density contains up to first derivatives of the scalar field) leading to second-order field equations, avoiding the appearance of Ostrogradsky instabilities and leading to ghost-free theory. These couplings, including linear and second-order curvature corrections, arise in the low energy effective action of  string theory (in fact a remarkable peculiarity of the string effective action is the appearance of field-dependent couplings to curvature) \cite{metsaev, cartier}, where couplings such as Gauss-Bonnet provide the possibility of avoiding the initial singularity \cite{antoniadis, Kanti:1998jd}. Given that there exist non-singular cosmological solutions based on these couplings, it is pertinent to investigate the effect of these correction terms on the evolution of primordial fluctuations that leave the power-spectrum nearly scale-invariant.
Also, in view of the accuracy of future observations, we expect that these corrections to the simplest, canonical scalar field, inflation model become important in a high-curvature regime typical of inflation. The effect of such corrections to the inflationary scenario could provide a connection with fundamental theories like supergravity or string theory. 
For studies of inflation with GB coupling and modified gravity see, for instance  \cite{Kanti:1998jd, soda4, Guo:2009uk, Guo:2010jr, Jiang:2013gza, Koh:2014bka, Kanti:2015pda, soda5, vandeBruck:2017voa, quiang, sergeioik, sergeioik1, chakraborty, zhuyi}. \\
In the appendixes we develop in detail the linear and quadratical perturbations for all the interaction terms of the model and deduce the second order action for the scalar and tensor perturbations. 
In appendix A we present the basic formulas for the first order perturbations, needed for the model, in the Newtonian gauge. In appendix B  we deduce the gravitational and scalar field equations in a general background. In appendix C and D we give the first order perturbations of the field equations in the Newtonian gauge. In Appendix E we give the details for constructing the second order action using the {\it Xpand} tool \cite{cyril}, and
in appendix F we give a detailed description of the slow-roll mechanism for the minimally coupled scalar field.\\
The expressions for the primordial density fluctuations in terms of the slow-roll parameters and the corresponding power spectra were found. We have found a consistency relation that is useful to discriminate the model from the standard inflation with canonical scalar field. The latest observational data disfavor monomial-type models $V\propto \phi^n$ with $n\ge 2$ in the minimally coupled scalar field. With the Introduction of additional interactions like the non-minimal coupling, kinetic coupling and Gauss-Bonnet coupling (GB), it is shown that the tensor-to-scalar ratio can be lowered to values that are consistent with latest observational constraints \cite{planck15, planck18}. This is sown in the case of quadratic potential with non-minimal and kinetic coupling, quadratic potential with kinetic and GB coupling and the general power-law potential with GB coupling. \\ 
The paper is organized as follows. In the next section we introduce the model, the background field equations and define the slow-roll parameters. In section 3 we use quadratic action for the scalar and tensor perturbations (details are given in the appendix) to evaluate the primordial power spectra. In section 4 we work some explicit models. Some discussion is given in section 5.
\section{The model and background equations}
We consider the scalar-tensor model with non-minimal coupling of the scalar field to curvature, non-minimal kinetic coupling of the scalar field to the Einstein's tensor and coupling of the scalar field to the Gauss-Bonnet (GB) 4-dimensional invariant
\be\label{eqm1}
S=\int d^4x\sqrt{-g}\left[\frac{1}{2}F(\phi)R-\frac{1}{2}\partial_{\mu}\phi\partial^{\mu}\phi-V(\phi)+F_1(\phi)G_{\mu\nu}\partial^{\mu}\phi\partial^{\nu}\phi-F_2(\phi){\cal G}\right]
\ee
where $G_{\mu\nu}$ is the Einstein's tensor, ${\cal G}$ is the GB 4-dimensional invariant given by
\be\label{eqm2}
{\cal G}=R^2-4R_{\mu\nu}R^{\mu\nu}+R_{\mu\nu\lambda\rho}R^{\mu\nu\lambda\rho}
\ee
\be\label{eqm3}
F(\phi)=\frac{1}{\kappa^2}+f(\phi),
\ee
and $\kappa^2=M_p^{-2}=8\pi G$.
One remarkable characteristic of this model is that it yields second-order field equations and can avoid Ostrogradski instabilities. Using the general results of Appendix B, expanded on the flat FRW background 
\be\label{eqm4}
ds^2=-dt^2+a(t)^2\left(dx^2+dy^2+dz^2\right)
\ee
one finds the following equations
\be\label{eqm5}
3H^2F\left(1-\frac{3F_1\dot{\phi}^2}{F}-\frac{8H\dot{F_2}}{F}\right)=\frac{1}{2}\dot{\phi}^2+V-3H\dot{F}
\ee
\be\label{eqm6}
\begin{aligned}
2\dot{H}F\left(1-\frac{F_1\dot{\phi}^2}{F}-\frac{8H\dot{F_2}}{F}\right)&=-\dot{\phi}^2-\ddot{F}+H\dot{F}+8H^2\ddot{F_2}-8H^3\dot{F_2}\\
&-6H^2F_1\dot{\phi}^2+4HF_1\dot{\phi}\ddot{\phi}+2H\dot{F}_1\dot{\phi}^2
\end{aligned}
\ee
\be\label{eqm7}
\begin{aligned}
\ddot{\phi}+3H\dot{\phi}&+V'-3F'\left(2H^2+\dot{H}\right)+24H^2\left(H^2+\dot{H}\right)F_2'+18H^3F_1\dot{\phi}\\&+12H\dot{H}F_1\dot{\phi}+6H^2F_1\ddot{\phi}+3H^2F_1'\dot{\phi}^2=0
\end{aligned}
\ee
where ($'$) denotes derivative with respect to the scalar field. Related to the different terms in the action (\ref{eqm1}) we define the following slow-roll parameters 
\be\label{eqm8}
\epsilon_0=-\frac{\dot{H}}{H^2},\;\;\; \epsilon_1=\frac{\dot{\epsilon}_0}{H\epsilon_0}
\ee
\be\label{eqm9}
\ell_0=\frac{\dot{F}}{HF},\;\;\; \ell_1=\frac{\dot{\ell}_0}{H\ell_0}
\ee
\be\label{eqm10}
k_0=\frac{3F_1\dot{\phi}^2}{F},\;\;\; k_1=\frac{\dot{k}_0}{Hk_0}
\ee
\be\label{eqm11}
\Delta_0=\frac{8H\dot{F_2}}{F},\;\;\; \Delta_1=\frac{\dot{\Delta}_0}{H\Delta_0}
\ee
The slow-roll conditions in this model are satisfied if all these parameters are much smaller than one, and will be used in the next section.
From the cosmological equations (\ref{eqm5}) and (\ref{eqm6}) and using the parameters (\ref{eqm8})-(\ref{eqm11}) we can write the following expressions for $\dot{\phi}^2$ and $V$
\be\label{eqm12}
\begin{aligned}
V=&H^2F\Big[3-\frac{5}{2}\Delta_0-2k_0-\epsilon_0+\frac{5}{2}\ell_0+\frac{1}{2}\ell_0\left(\ell_1-\epsilon_0+\ell_0\right)\\&-\frac{1}{2}\Delta_0\left(\Delta_1-\epsilon_0+\ell_0\right)-\frac{1}{3}k_0\left(k_1+\ell_0-\epsilon_0\right)\Big]
\end{aligned}
\ee
\be\label{eqm13}
\begin{aligned}
\dot{\phi}^2=&H^2F\Big[2\epsilon_0+\ell_0-\Delta_0-2k_0+\Delta_0\left(\Delta_1-\epsilon_0+\ell_0\right)-\\& \ell_0\left(\ell_1-\epsilon_0+\ell_0\right)+\frac{2}{3}k_0\left(k_1+\ell_0-\epsilon_0\right)\Big]
\end{aligned}
\ee
where we used
\be\label{eqm13a}
\ddot{F}=H^2F\ell_0\left(\ell_1-\epsilon_0+\ell_0\right),\;\;\; \ddot{F}_2=\frac{F\Delta_0}{8}\left(\Delta_1+\epsilon_0+\ell_0\right)
\ee
It is also useful to define the variable $Y$ from Eq. (\ref{eqm13}) as
\be\label{eqm13b}
Y=\frac{\dot{\phi}^2}{H^2F}
\ee
where it follows that $Y = {\cal O}(\varepsilon )$. Notice that for the simplest case of minimally coupled scalar field ($F=1/\kappa^2$, $F_1=F_2=0$), the Eqs. (\ref{eqm12}) and (\ref{eqm13}) give the standard equations
\be\nonumber
H^2=\frac{8\pi G}{3}\left(\frac{1}{2}\dot{\phi}^2+V(\phi)\right),\;\;\; \dot{H}=-4\pi G\dot{\phi}^2
\ee
Under the slow-roll conditions $\ddot{\phi}<<3H\dot{\phi}$ and $\ell_i, k_i, \Delta_i<<1$, it follows from (\ref{eqm5})-(\ref{eqm7}) 
\be\label{eqm14}
3H^2F\simeq V,
\ee
\be\label{eqm15}
2\dot{H}F\simeq -\dot{\phi}^2+H\dot{F}-6H^2F_1\dot{\phi}^2-8H^3\dot{F}_2,
\ee
\be\label{eqm16}
3H\dot{\phi}+V'-6H^2F'+18H^3F_1\dot{\phi}+24H^4F'_2\simeq 0
\ee
showing that the potential $V$ gives the dominant contribution to the Hubble parameter, while Eqs. (\ref{eqm15}) and (\ref{eqm16}) determine the dynamics of the scalar field in the slow-roll approximation. The number of $e$-folds can be determined from
\be\label{eqm17}
N=\int_{\phi_I}^{\phi_E}\frac{H}{\dot{\phi}}d\phi=\int_{\phi_I}^{\phi_E}\frac{H^2+6H^4F_1}{2H^2F'-8H^4F'_2-\frac{1}{3}V'}d\phi
\ee
where $\phi_I$ and $\phi_E$ are the values of the scalar field at the beginning and end of inflation respectively, and the expression for $\dot{\phi}$ was taken from (\ref{eqm16}). The criteria for choosing the initial values will be discussed below.
\section{Quadratic action for the scalar and tensor perturbations}
{\bf Scalar Perturbations}.\\

\noindent After the computation of the second order perturbations we are able to write the second order action for the scalar perturbations as follows
\be\label{slr1}
\delta S_s^{2}=\int dt d^3xa^3\left[{\cal G}_s\dot{\xi}^2-\frac{{\cal F}_s}{a^2}\left(\nabla\xi\right)^2\right]
\ee
where
\be\label{slr2}
{\cal G}_s=\frac{\Sigma}{\Theta^2}{\cal G}_T^2+3{\cal G}_T
\ee
\be\label{slr3}
{\cal F}_s=\frac{1}{a}\frac{d}{dt}\left(\frac{a}{\Theta}{\cal G}_T^2\right)-{\cal F}_T
\ee
with
\be\label{slr4}
{\cal G}_T=F-F_1\dot{\phi}^2-8H\dot{F}_2.
\ee
\be\label{slr5}
{\cal F}_T=F+F_1\dot{\phi}^2-8\ddot{F}_2
\ee
\be\label{slr6}
\Theta=FH+\frac{1}{2}\dot{F}-3HF_1\dot{\phi}^2-12H^2\dot{F}_2
\ee
\be\label{slr7}
\Sigma=-3FH^2-3H\dot{F}+\frac{1}{2}\dot{\phi}^2+18H^2F_1\dot{\phi}^2+48H^3\dot {F}_2
\ee
And the sound speed of scalar perturbations is given by 
\be\label{slr8}
c_S^2=\frac{{\cal F}_S}{{\cal G}_S}
\ee
The conditions for avoidance of ghost and Laplacian instabilities as seen from the action (\ref{slr1}) are
$$ {\cal F}>0,\;\;\;\; {\cal G}>0 $$
We can rewrite ${\cal G}_T$, ${\cal F}_T$, $\Theta$ and $\Sigma$ in terms of the slow-roll parameters (\ref{eqm8})-(\ref{eqm11}) and using Eqs. (\ref{eqm13}) and (\ref{eqm13a}), as follows
\be\label{slr8a}
{\cal G}_T=F\left(1-\frac{1}{3}k_0-\Delta_0\right)
\ee
\be\label{slr8b}
{\cal F}_T=F\left(1+\frac{1}{3}k_0-\Delta_0\left(\Delta_1+\epsilon_0+\ell_0\right)\right)
\ee
\be\label{slr8c}
\Theta=FH\left(1+\frac{1}{2}\ell_0-k_0-\frac{3}{2}\Delta_0\right)
\ee
\be\label{slr8d}
\begin{aligned}
\Sigma=&-FH^2\Big[3-\epsilon_0+\frac{5}{2}\ell_0-5k_0-\frac{11}{2}\Delta_0+\frac{1}{2}\ell_0\left(\ell_1-\epsilon_0+\ell_0\right)\\&-\frac{1}{3}k_0\left(k_1-\epsilon_0+\ell_0\right)-\frac{1}{2}\Delta_0\left(\Delta_1-\epsilon_0+\ell_0\right)\Big]
\end{aligned}
\ee
The expressions for ${\cal G}_S$ and $c_S^2$ in terms of the slow roll parameters can be written as
\be\label{slowGS}
{{\cal G}_S} = \frac{{F\left( {\frac{1}{2}Y + {k_0} + \frac{3}{4}{W^2}(1 - {\Delta _0} - \frac{1}{3}{k_0})} \right)}}{{{{\left( {1 + \frac{1}{2}W} \right)}^2}}}
\ee
\be\label{slowCS}
c_S^2 = 1 + \frac{{{W^2}\left( {\frac{1}{2}{\Delta _0}({\Delta _1} + {\varepsilon _0} + {l_0} - 1) - \frac{1}{3}{k_0}} \right) + W\left( {\frac{2}{3}{k_0}\left( {2 - {k_1} - {l_0}} \right) + 2{\Delta _0}{\varepsilon _0}} \right) - \frac{4}{3}{k_0}{\varepsilon _0}}}{{Y + 2{k_0} + \frac{3}{2}{W^2}(1 - {\Delta _0} - \frac{1}{3}{k_0})}}
\ee
where

\be\label{slowW}
W = \frac{\ell_0-\Delta_0-\frac{4}{3}k_0}{1-\Delta_0-\frac{1}{3}k_0}
\ee
Notice that in general ${\cal G}_S=F {\cal O}(\varepsilon )$ and $c_S^2 = 1 +{\cal O}(\varepsilon )$. Also in absence of the kinetic coupling it follows that $c_S^2 = 1 +{\cal O}(\varepsilon^2 )$. Keeping first order terms in slow-roll parameters, the expressions for ${\cal G}_S$ y $c_S^2$ reduce to 
\be\label{aproxGS}
{G_S} = F\left( {{\varepsilon _0} + \frac{1}{2}{l_0} - \frac{1}{2}{\Delta _0}} \right)
\ee
\be\label{aproxCS}
c_S^2 = 1 + \frac{{\frac{4}{3}{k_0}\left( {{l_0} - {\Delta _0} - \frac{4}{3}{k_0}} \right) - \frac{4}{3}{k_0}{\varepsilon _0}}}{{2{\varepsilon _0} + {l_0} - {\Delta _0}}}
\ee 
\noindent To normalize the scalar perturbations we perform the change of variables \cite{kobayashi1} (see (\ref{f11}))
\be\label{slr9}
d\tau_s=\frac{c_S}{a}dt,\;\;\, \tilde{z}=\sqrt{2}a\left({\cal F}_S{\cal G}_S\right)^{1/4},\;\;\; \tilde{U}=\xi\tilde{z}
\ee
and the action (\ref{slr1}) becomes 
\be\label{slr10}
\delta S_s^{2}=\frac{1}{2}\int d\tau_s d^3x\left[\frac{1}{2}(\tilde{U}')^2-D_i\tilde{U}D^{i}\tilde{U}+\frac{\tilde{z}''}{\tilde{z}}\tilde{U}^2\right]
\ee
where  "prima" indicates derivative with respect to $\tau_s$. Working in the Fourier representation, one can write 
\be\label{slr11}
\tilde{U}(\vec{x},\tau_s)=\int\frac{d^3k}{(2\pi)^3}\tilde{U}_{\vec{k}}(\tau_s)e^{i\vec{k}\vec{x}}
\ee
and the equation of motion for the action (\ref{slr10}) takes the form
\be\label{slr12}
\tilde{U}_{\vec{k}}''+\left(k^2-\frac{\tilde{z''}}{\tilde{z}}\right)\tilde{U}_{\vec{k}}=0
\ee
From (\ref{slr9}), and keeping up to first-order terms in slow-roll variables using (\ref{aproxGS}) and (\ref{aproxCS}), we find the following expression for $\tilde{z}'$  
\be\label{slr13}
\tilde{z}'=\frac{1}{c_S}\frac{a^5}{z^3}\left[F^2\frac{df(\epsilon_0,\ell_0,\Delta_0)}{dt}+2F\dot{F}f(\epsilon_0,\ell_0,\Delta_0)\right]+\frac{1}{c_S}aHz
\ee
where
$$
f(\epsilon_0,\ell_0,\Delta_0)=\left(\epsilon_0+\frac{1}{2}\ell_0-\frac{1}{2}\Delta_0\right)^2.
$$
Then, under the approximation of slowly varying $c_S$ and up to first-order in slow-roll variables we find the following expression for $\tilde{z}''/\tilde{z}$
\be\label{slr14}
\frac{\tilde{z}''}{\tilde{z}}=\frac{a^2H^2}{c_S^2}\Big[2-\epsilon_0+\frac{3}{2}\ell_0+\frac{3}{2}\frac{2\epsilon_0\epsilon_1+\ell_0\ell_1-\Delta_0\Delta_1}{2\epsilon_0+\ell_0-\Delta_0}\Big].
\ee
This expression reduces to the one of the canonical scalar field given in Appendix E, Eq. (\ref{f25}), in the case $\ell_0=\Delta_0=0$ where $c_S=1$ and $\epsilon_1=2(\epsilon_0-\delta)$, with $\delta$ defined in (\ref{f23}). In what follows the reasoning is similar to the simplest case, corresponding to minimally-coupled scalar field, which is analyzed in detail in Appendix E. So on sub-horizon scales when the $k^2$ term dominates in Eq. (\ref{slr12}) we choose the same Bunch-Davies vacuum solution defined for the scalar field, which leads to
\be\label{slr15}
\tilde{U}_k=\frac{1}{\sqrt{2k}}e^{-ik\tau_s}
\ee
Note that from the expression 
\be\label{slr16}
a\frac{d}{dt}\left(\frac{1}{aH}\right)=-1+\epsilon\Longrightarrow c_S\frac{d}{d\tau_s}\left(\frac{1}{aH}\right)=-1+\epsilon_0,
\ee
in the approximation of slowly varying $c_S$ and $\epsilon_0$ one can integrate the last equation to obtain
\be\label{slr17}
\tau_s=-\frac{1}{aH}\frac{c_S}{1-\epsilon_0}
\ee
Then in the limit $\epsilon\rightarrow 0$ for de Sitter expansion it follows that
\be\label{slr18}
\frac{1}{aH}=-\frac{\tau_{dS}}{c_S}
\ee 
In this last case and neglecting the slow-roll parameters (in this limit $c_S=1$) we can write from (\ref{slr14})
\be\label{slr19}
\frac{\tilde{z}''}{\tilde{z}}\simeq \frac{2a^2H^2}{c_S^2}=\frac{2}{\tau_{dS}^2}
\ee
which allows the integration of Eq. (\ref{slr12}), giving the known solution for the scalar perturbations in a de Sitter background. 
Taking into account the slow-roll parameters and using (\ref{slr17}) we can rewrite the Eq. (\ref{slr12}) in the form
\be\label{slr20}
\tilde{U}_k''+k^2\tilde{U}_{k}+\frac{1}{\tau_s^2}\left(\mu_s^2-\frac{1}{4}\right)\tilde{U}_{k}=0
\ee
where
\be\label{slr21}
\mu_s^2=\frac{9}{4}\left[1+\frac{4}{3}\epsilon_0+\frac{2}{3}\ell_0+\frac{2}{3}\frac{2\epsilon_0\epsilon_1+\ell_0\ell_1-\Delta_0\Delta_1}{2\epsilon_0+\ell_0-\Delta_0} \right]
\ee
where we have expanded up to first order in slow-roll parameters. The general solution of Eq. (\ref{slr21}) for constant $\mu_s$ (slowly varying slow-roll parameters) is 
\be\label{slr22}
\tilde{U}_k=\sqrt{-\tau_s}\Big[C_{1k}H_{\mu_s}^{(1)}(-k\tau_s)+C_{2k}H_{\mu_s}^{(2)}(-k\tau_s)\Big]
\ee
and after matching the boundary condition related with the choosing of the Bunch-Davies vacuum (\ref{slr15}) we find the solution
\be\label{slr23}
\tilde{U}_k=\frac{\sqrt{\pi}}{2}e^{i\frac{\pi}{2}(\mu_s+\frac{1}{2})}\sqrt{-\tau_s}H_{\mu_s}^{(1)}(-k\tau_s)
\ee
using the asymptotic behavior of $H_{\mu_s}^{(1)}(x)$ at $x>>1$, we find at super horizon scales ($c_Sk<<aH$)
\be\label{slr24}
\tilde{U}_k=\frac{1}{\sqrt{2}}e^{i\frac{\pi}{2}(\mu_s-\frac{1}{2})}2^{\mu_s-\frac{3}{2}}\frac{\Gamma(\mu_s)}{\Gamma(3/2)}\sqrt{-\tau_s}(-k\tau_s)^{-\mu_s}.
\ee
To evaluate the power spectra we use the relationship
\be\label{slr25}
\frac{\tilde{z}'}{\tilde{z}}=-\frac{1}{(1-\epsilon_0)\tau_s}\Big[1+\frac{1}{2}\ell_0+\frac{1}{2}\frac{2\epsilon_0\epsilon_1+\ell_0\ell_1-\Delta_0\Delta_1}{2\epsilon_0+\ell_0-\Delta_0}\Big]=-\frac{1}{\tau_s}\left(\mu_s-\frac{1}{2}\right).
\ee
where we used (\ref{slr17}) for $aH$, and for the last equality we have expanded up to first order in slow-roll parameters, resulting in 
\be\label{slr25a}
\mu_s=\frac{3}{2}+\epsilon_0+\frac{1}{2}\ell_0+\frac{1}{2}\frac{2\epsilon_0\epsilon_1+\ell_0\ell_1-\Delta_0\Delta_1}{2\epsilon_0+\ell_0-\Delta_0}
\ee
Assuming again the approximation of slowly varying slow-roll parameters we can Integrate this equation to find
\be\label{slr26}
\tilde{z}\propto \tau_s^{\frac{1}{2}-\mu_s}
\ee
which gives, in the super horizon regime, for the amplitude of the scalar perturbations the following expression
\be\label{slr27}
\xi_k=\frac{\tilde{U}_k}{\tilde{z}}\propto k^{-\mu_s}
\ee
where the $\tau_s$ dependence disappears as expected from the solution (\ref{slr24}) in super horizon scales ($c_sk<<aH$). The power spectra for the scalar perturbations takes the following $k$-dependence
\be\label{slr28}
P_{\xi}=\frac{k^3}{2\pi^2}|\xi_k|^2\propto k^{3-2\mu_s}
\ee
and the scalar spectral index becomes
\be\label{slr29}
n_s-1=\frac{d\ln P_{\xi}}{d\ln k}=3-2\mu_s=-2\epsilon_0-\ell_0-\frac{2\epsilon_0\epsilon_1+\ell_0\ell_1-\Delta_0\Delta_1}{2\epsilon_0+\ell_0-\Delta_0}
\ee
It is worth noticing that the slow-roll parameter $k_0$, related to the kinetic coupling, do not appear in the above expression for the scalar spectral index. This is because $k_0$ appears only in second order terms (or higher) in the expressions for ${\cal G}_S$ and ${\cal F}_S$ (see (\ref{slowGS}) and (\ref{slowCS})).\\

\noindent {\bf Tensor perturbations}.\\

\noindent The second order action for the tensor perturbations takes the form
\be\label{slrt1}
\delta S_2=\frac{1}{8}\int d^3xdt{\cal G}_T a^2\left[\left(\dot{h}_{ij}\right)^2-\frac{c_T^2}{a^2}\left(\nabla h_{ij}\right)^2\right]
\ee 
where ${\cal G}_T$ and ${\cal F}_T$ are defined in (\ref{slr4}) and (\ref{slr5}) (in terms of the slow-roll variables in (\ref{slr9}) and (\ref{slr10})). The velocity of tensor perturbations is given by
\be\label{slrt2}
c_T^2=\frac{{\cal F}_T}{{\cal G}_T}=\frac{3+k_0-3\Delta_0\left(\Delta_1+\epsilon_0+\ell_0\right)}{3-k_0-3\Delta_0}.
\ee
As in the case of scalar perturbations, in order to canonically normalize the tensor perturbations the following variables are used \cite{kobayashi1}
\be\label{slrt3}
d\tau_T=\frac{c_T}{a}dt,\;\;\; z_T=\frac{a}{2}\left({\cal F}_T{\cal G}_T\right)^{1/4},\;\;\; v_{ij}=z_T h_{ij}
\ee
leading to the quadratic action
\be\label{slrt4}
\delta S_2=\frac{1}{2}\int d^3xd\tau_T\left[\left(v'_{ij}\right)^2-\left(\nabla v_{ij}\right)^2+\frac{z''_T}{z_T}v_{ij}^2\right]
\ee
which gives the equation
\be\label{slrt5}
v''_{ij}-\nabla^2 v_{ij}-\frac{z''_T}{z_T}v_{ij}=0.
\ee
Or for the corresponding Fourier modes
\be\label{slrt6}
v''_{(k)ij}+\left(k^2-\frac{z''_T}{z_T}\right)v_{(k)ij}=0,
\ee
which is of the same nature as the equation for the scalar perturbations, and therefore the perturbations $h_{ij}$ on super horizon scales behave exactly as the solutions (\ref{f10}). For the evaluation of the primordial power spectrum we follow the same steps as for the scalar perturbations. 
To this end we write the expression for $z''_T/z_T$, up to first order in slow-roll parameters, as follows
\be\label{slr7}
\frac{z''_T}{z_T}=\frac{a^2H^2}{c_T^2}\left(2-\epsilon_0+\frac{3}{2}\ell_0\right)
\ee
Then, the normalized solution of (\ref{slrt6}) in the approximation of slowly varying slow-roll parameters can be written in terms of the Hankel function of the first kind as
\be\label{slrt9}
v_{(k)ij}=\frac{\sqrt{\pi}}{2}\sqrt{-\tau_T}H_{\mu_T}^{(1)}(-k\tau_T)e^{(k)}_{ij}
\ee
where the tensor $e^{(k)}_{ij}$ describe the polarization states of the tensor perturbations for the $k$-mode, and 
\be\label{slrt10}
\mu_T=\frac{3}{2}+\epsilon_0+\frac{1}{2}\ell_0.
\ee
At super horizon scales ($c_Tk<<aH$) the tensor modes (\ref{slrt9}) have the same functional form for the asymptotic behavior as the scalar modes (\ref{slr24}), and therefore we can write power spectrum for tensor perturbations as
\be\label{slrt11}
P_T=\frac{k^3}{2\pi^2}|h^{(k)}_{ij}|^2
\ee
where $h^{(k)}_{ij}=v_{(k)ij}/z_T$, and the sum over the polarization states must be taken into account. 
Then, the tensor spectral index will be given by
\be\label{slrt11}
n_T=3-2\mu_T=-2\epsilon_0-\ell_0
\ee
An important quantity is the relative contribution to the power spectra of tensor and scalar perturbations, defined as the tensor/scalar ratio $r$
\be\label{slrt12}
r=\frac{P_T(k)}{P_{\xi}(k)}.
\ee
For the scalar perturbations, using (\ref{slr28}), we can write the power spectra as
\be\label{slrt13}
P_{\xi}=A_S\frac{H^2}{(2\pi)^2}\frac{{\cal G}_S^{1/2}}{{\cal F}_S^{3/2}}
\ee
where 
$$A_S=\frac{1}{2}2^{2\mu_s-3}\Big|\frac{\Gamma(\mu_s)}{\Gamma(3/2)}\Big|^2$$
and all magnitudes are evaluated at the moment of horizon exit when $c_s k=aH$ ($k\tau_s=-1$). For $\tilde{z}$ we used (\ref{slr9}) with $a=c_S k/H$. In analogous way we can write the power spectra for tensor perturbations as
\be\label{slrt14}
P_T= 16A_T\frac{H^2}{(2\pi)^2}\frac{{\cal G}_T^{1/2}}{{\cal F}_T^{3/2}}
\ee
where 
$$A_T=\frac{1}{2}2^{2\mu_T-3}\Big|\frac{\Gamma(\mu_T)}{\Gamma(3/2)}\Big|^2.$$
Noticing that $A_T/A_S\simeq 1$ when evaluated at the limit $\epsilon_0,\ell_0,\Delta_0,...<<1$, as follows from (\ref{slr25a}) and (\ref{slrt10}), we can write the tensor/scalar ratio as follows
\be\label{slrt15}
r=16\frac{{\cal G}_T^{1/2}{\cal F}_S^{3/2}}{{\cal G}_S^{1/2}{\cal F}_T^{3/2}}=16\frac{c_S^3{\cal G}_S}{c_T^3{\cal G}_T}.
\ee
Taking into account the expressions for ${\cal G}_T, {\cal F}_T, {\cal G}_S,{\cal F}_S$ up to first order obtained from (\ref{slr8a}), (\ref{slr8b}), (\ref{aproxGS}) and (\ref{aproxCS}), and using the condition $\epsilon_0,\ell_0,k_0,\Delta_0<<1$, then  we can see that $c_T\simeq c_S\simeq 1$ (in fact in the limit $\ell_0\rightarrow 0$, $c_S=1$ independently of the values of $\epsilon_0$ and $\Delta_0$) and we can make the approximation
\be\label{slrt16}
r=8\left(\frac{2\epsilon_0+\ell_0-\Delta_0}{1-\frac{1}{3}k_0-\Delta_0}\right)\simeq 8\left(2\epsilon_0+\ell_0-\Delta_0\right)
\ee
which is a modified consistency relation due to the non-minimal and GB couplings. In the limit $\ell_0,\Delta_0\rightarrow 0$ it gives the expected  consistency relation for the standard inflation
\be\label{slrt17}
r= -8n_T,
\ee
with $n_T=-2\epsilon_0$. Taking into account the non-minimal and GB couplings we find the deviation from the standard consistency relation in the form
\be\label{slrt18}
r=-8n_T+\delta r,\;\;\; \delta r=-8\Delta_0,
\ee
with $n_T$ given by (\ref{slrt11}). Thus, the consistency relation still valid in the case of non-minimal coupling, and if there is an observable appreciable deviation from the standard consistency relation, it can reveal the effect of an interaction beyond the simple canonical scalar field
or even non-minimally coupled scalar field models of inflation.
It is worth noticing that in the first-order formalism the kinetic-coupling related slow-roll parameter $k$ does not appear in the spectral index for the scalar and tensor perturbations and is also absent in the tensor-to-scalar ratio, appearing only starting form the second order expansion in slow-roll parameters. Nevertheless, all the couplings are involved in the definition of the slow-roll parameters trough the field equations. Of special interest are the cases of monomial potentials $V\propto \phi^n$. These potentials are disfavored by the observational data for $n\ge 2$ in the minimally coupled model. As will be shown for some cases, with the GB and (or) kinetic coupling added, the spectral index and especially the scalar-to tensor ratio can be accommodated within the range of values obtained from the latest observational data.
\section{Some explicit cases}
{\bf Model I.}\\
First we consider the particular case of the non-minimal coupling $\xi\phi^2$ with quadratic potential and kinetic coupling with constant $F_1$. 
\be\label{ejea1}
F\left( \phi  \right) = \frac{1}{{{\kappa ^2}}} - \xi {\phi ^2},\,\,\,\,\,\,\,\,\,V\left( \phi  \right) = \frac{{{1}}}{{{2}}} m^2\phi^2,\,\,\,\,\,\,\,\,\,{F_1}\left( \phi  \right) = \gamma ,\,\,\,\,\,\,\,\,{F_2}\left( \phi  \right) = 0.
\ee
Using the Eqs. (\ref{eqm14}) and (\ref{eqm16}) we can express the slow-roll parameters (\ref{eqm8})-(\ref{eqm11}) in therms of the potential and the coupling functions, and once we specify the model, we can find the slow-roll parameters in terms of the scalar field and the coupling constants. For the model (\ref{ejea1}) the slow-roll parameters take the form
\be\nonumber
{\epsilon _0} = \frac{2 + 2\xi \phi ^2}{\phi ^2 + (m^2\gamma-\xi) \phi ^4},\,\,\,\,\,\,\,\,\,\,\,\,\,\,\,{\epsilon _1} = \frac{4(1-\xi\phi^2)\left((m^2\gamma-\xi)(\xi\phi^2+2)\phi^2+1\right)}{\phi^2\left(1+(m^2\gamma-\xi)\phi^2\right)^2}
\ee
\be\label{ejea2}
\ell{_0} =  \frac{4\xi(\xi\phi^2+1)}{(m^2\gamma-\xi)\phi^2+1},\,\,\,\,\,\,\,\,\,\,\,\,{\ell_1} = -\frac{4(m^2\gamma-2\xi)(\xi\phi^2-1)}{((m^2\gamma-\xi)\phi^2+1)^2}.
\ee
where $\phi$ is dimensionless ($\phi$ has been rescaled as $\kappa\phi\rightarrow\phi$ to measure it in units of $M_p$) and $\gamma$ has dimension of $mass^{-2}$ . Additionally, the scalar field at the end of inflation can be evaluated under the condition ${\epsilon _0}({\phi _E})=1$. 
Sitting ${\epsilon _0}=1$ in (\ref{ejea2}) it follows
\be\label{ejea3}
\phi_E^2=\frac{\sqrt{8m^2\gamma+4\xi^2-12\xi+1}+2\xi-1}{2m^2\gamma-2\xi}
\ee
From Eq. \ref{eqm17} it follows that the number of $e$-foldings can be evaluated as
\be\label{ejea4}
N = \int\limits_{{\phi _I}}^{{\phi _E}} { \frac{{\phi  + (m^2\gamma-\xi) {\phi ^3}}}{{2\xi^2 {\phi ^4}}-2}d\phi }=\frac{1}{8\xi^2}\left[m^2\gamma\ln\left(1-\xi^2\phi^4\right)-2\xi\ln\left(1-\xi\phi^2\right)\right] \Big|_{\phi_I}^{\phi_E}
\ee
This expression allows us to evaluate $\phi _I$ for a given $N$. We can make some qualitative analysis by assuming that $\xi\phi^2<<1$ and $m^2\gamma>>\xi$. In this case from (\ref{ejea3}) it is found that
\be\label{ejea5}
\phi_E^2\approx \left(\frac{2}{m^2\gamma}\right)^{1/2},
\ee
and from (\ref{ejea4}) we find for $\phi_I$
\be\label{ejea6}
\phi_I^2\approx \left(\frac{8N+2}{m^2\gamma}\right)^{1/2} 
\ee
giving an approximate relation between the values of the scalar field at the beginning and end of inflation as 
$$\phi_I\approx (4N+1)^{1/4}\phi_E$$
So, assuming $N=60$ gives $\phi_I\approx 3.9\phi_E$. This will have sense only if the scalar spectral index and the tensor-to-scalar ratio behave properly. In fact from (\ref{ejea2}) and replacing in (\ref{slr29}) and (\ref{slrt16}), we find (under the condition $\xi\phi^2<<1$ and $m^2\gamma>>\xi$)
\be\label{ejea7}
n_s\approx 1+\frac{2}{(8N+2)^{1/2}(m^2\gamma)^{1/2}}-\frac{12}{8N+2}-\frac{8}{(8N+2)^{3/2}(m^2\gamma)^{1/2}}
\ee
and
\be\label{ejea8}
r\approx \frac{32}{8N+2}+\frac{64\xi}{(8N+2)^{1/2}(m^2\gamma)^{1/2}}
\ee
where we have used (\ref{ejea6}) for $\phi_I$. Additional simplification can be made if we assume that the scalar field at the beginning of inflation is of the order of $M_p$ ($\phi\simeq 1$). This can be achieved if $m^2\gamma=8N+2$, as follows from (\ref{ejea6}), which gives
\be\label{ejea9}
n_s\approx 1-\frac{10}{8N+2}-\frac{8}{(8N+2)^2},\;\;\;\; r\approx \frac{32+64\xi}{8N+2}
\ee 
Thus, for 60 $e$-foldings we find $n_s\approx 0.98$ and $r\approx 0.067$ ($\xi=10^{-2}$). In this case the inflation begins with $\phi_I=M_p$ and ends with $\phi_E\approx 0.25M_p$. 
For the numerical analysis with the exact expressions, we assume $N=60$, $m=10^{-6}M_p$. In fact from Eqs. (\ref{ejea2}) follows that the spectral index $n_s$ and the tensor-to-scalar ratio depend on the dimensionless combination $m^2\gamma$. Fig. 1 shows the behavior of $n_s$ and $r$ where $\xi$ takes two fixed values $\xi=1/6,\; \xi= 0.2$ and  $m^2\gamma$ is running in the interval $2\times 10^2\le m^2\gamma\le 5\times 10^{2}$, and in Fig. 2 we consider the two fixed values $m^2\gamma=1$ and $m^2\gamma=2.5$ while $\xi$ is varying in the interval $-0.04\le\xi\le -0.01$.\\
\begin{figure}[hbtp]
\centering
\includegraphics[scale=0.7]{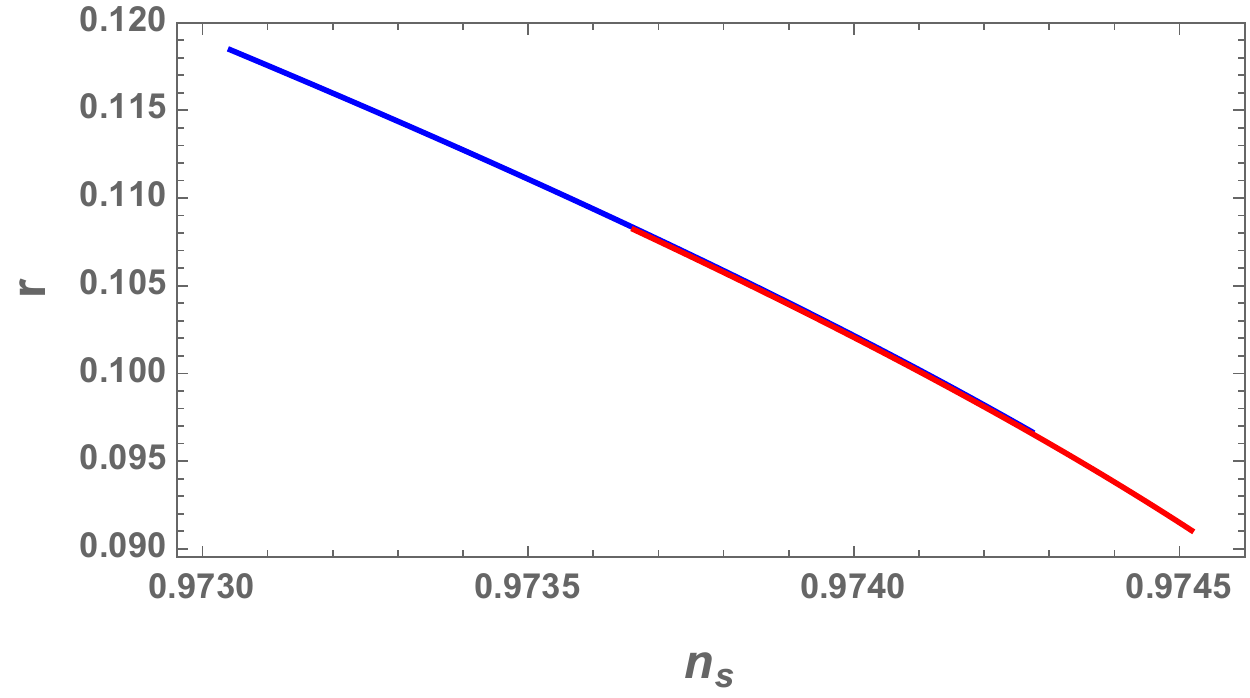}
\caption{ The behavior of the scalar spectral index $n_s$ and tensor-to-scalar ratio $r$ for the two fixed values $\xi=1/6$ (red) and $\xi=0.2$ (blue), with $m^2\gamma$ in the interval $2\times 10^2\le m^2\gamma\le 5\times 10^{2}$.}
\end{figure}
\begin{figure}[hbtp]
\centering
\includegraphics[scale=0.7]{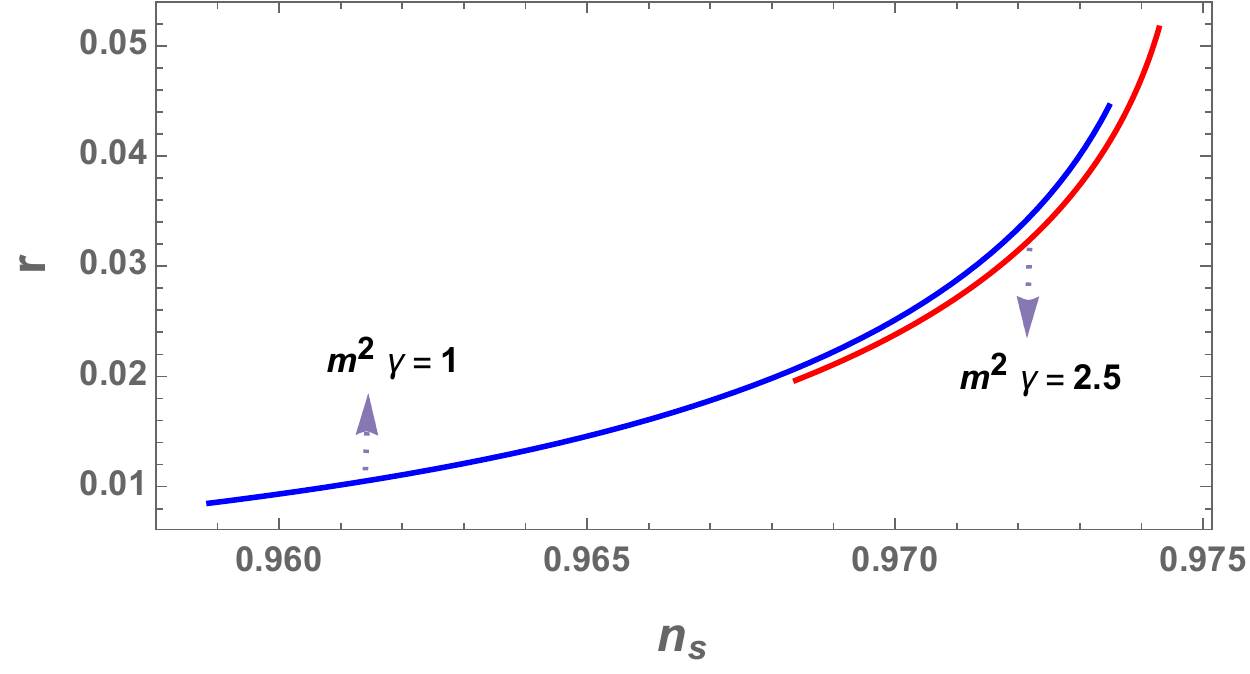}
\caption{The variation of the scalar spectral index $n_s$ and tensor-to-scalar ratio $r$ for the two cases $m^2\gamma=1$ and $m^2\gamma=2.5$, for $\xi$ in the interval $[-0.04,-0.01]$. The values covered by $n_s$ and $r$ are in the region bounded by the latest observations \cite{planck18, bkp}.}
\end{figure}
\noindent The behavior shown in Fig. 2 is more interesting than the one in Fig. 1, since the values for  $n_s$ and $r$ fall in a more acceptable range, according to the latest observational bounds \cite{planck18, bkp}. Thus, keeping $m^2\gamma\sim 1$ with $\xi$ varying in the interval $[-0.04,-0.01]$ gives the expected values for the observables, according to latest observations. \\

\noindent {\bf Model II}\\
\noindent The following example considers a model with kinetic and GB couplings 
\be\label{ejeb1}
F=\frac{1}{\kappa^2},\;\;\; V(\phi)=\frac{1}{2}m^2\phi^2,\;\;\; F_1(\phi)=\gamma,\;\;\; F_2(\phi)=\frac{\eta}{\phi^2}
\ee
where the constant $\eta$ has dimension of $mass^2$ and $\phi$ is measured in units of $M_p$. The slow-roll parameters from (\ref{eqm8})-(\ref{eqm11}), necessary to evaluate  $n_s$ and $r$, take the form
\be\nonumber
\epsilon_0=\frac{6-8m^2\eta}{3\phi^2(1+m^2\gamma\phi^2)},\;\;\; \epsilon_1=\frac{4(3-4m^2\eta)(1+2m^2\gamma\phi^2)}{3\phi^2(1+m^2\gamma\phi^2)^2}
\ee
\be\label{ejeb2}
\Delta_0=\frac{16m^2\eta(3-4m^2\eta)}{9\phi^2(1+m^2\gamma\phi^2)},\;\;\; \Delta_1=\frac{4(3-4m^2\eta)(1+2m^2\gamma\phi^2)}{3\phi^2(1+m^2\gamma\phi^2)^2},
\ee
where theproduct $m^2\eta$ is measured in units of $M_p^4$ and the product $m^2\gamma$ is dimensionless. The scalar field at the end of inflation is obtained from the condition $\epsilon(\phi_E)=1$, which gives
\be\label{ejeb3}
\phi_E^2=\frac{1}{6m^2\gamma}\left[\sqrt{72m^2\gamma-96m^4\gamma\eta+9}-3\right]
\ee
And From Eq. (\ref{eqm17}), the number of $e$-foldings can be evaluated as
\be\label{ejeb4}
N = \frac{3\phi^2(2+m^2\gamma\phi^2)}{8(4m^2\eta-3)} \Big|_{\phi_I}^{\phi_E}
\ee
which allows to find $\phi_I$ for a given $N$ and $\phi_E$ from (\ref{ejeb3}). From (\ref{slr29}) and (\ref{ejeb2}) we find the scalar spectral index as
\be\label{ejeb5}
n_s=\frac{3m^4\gamma\phi^2\left(\gamma\phi^4+16\eta\right)+6m^2\gamma\phi^2(\phi^2-6)+3\phi^2+32m^2\eta-24}{3\phi^2(1+m^2\gamma\phi^2)^2}\Big|_{\phi_I}
\ee
And from (\ref{slrt16}) and (\ref{ejeb2}) we find the expression for the tensor-to-scalar ratio as
\be\label{ejeb6}
r=\frac{3\phi(1+m^2\gamma\phi^2)}{8m^2\eta-6}\Big|_{\phi_I}
\ee
For $N=60$ and taking $m=10^{-6}M_p$ we can find the behavior of $n_s$ and $r$ in terms of the dimensionless parameter $m^2\gamma$. In Fig. 3 we show the behavior of the scalar field at the beginning and end of inflation for $1<m^2\gamma<5$. In Fig. 4 we show the corresponding behavior for $n_s$ and $r$.\\
\begin{figure}
\centering
\includegraphics[scale=0.7]{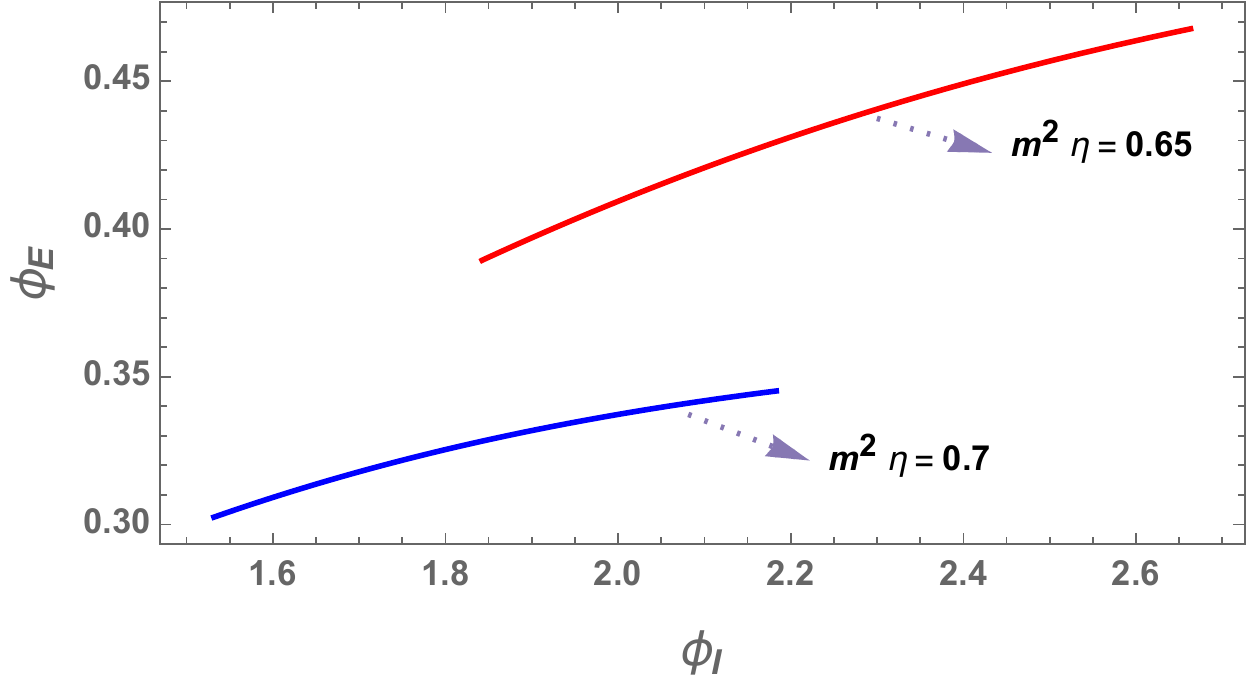}
\caption{The scalar field at the beginning and at the end of inflation for $1\le m^2\gamma\le 5$ and $m^2\eta=0,65$, $m^2\eta=0.7$ (in units of $M_p^4$). At the end of inflation $\phi_E<M_p$.}
\end{figure}

\noindent 
\begin{figure}
\centering
\includegraphics[scale=0.7]{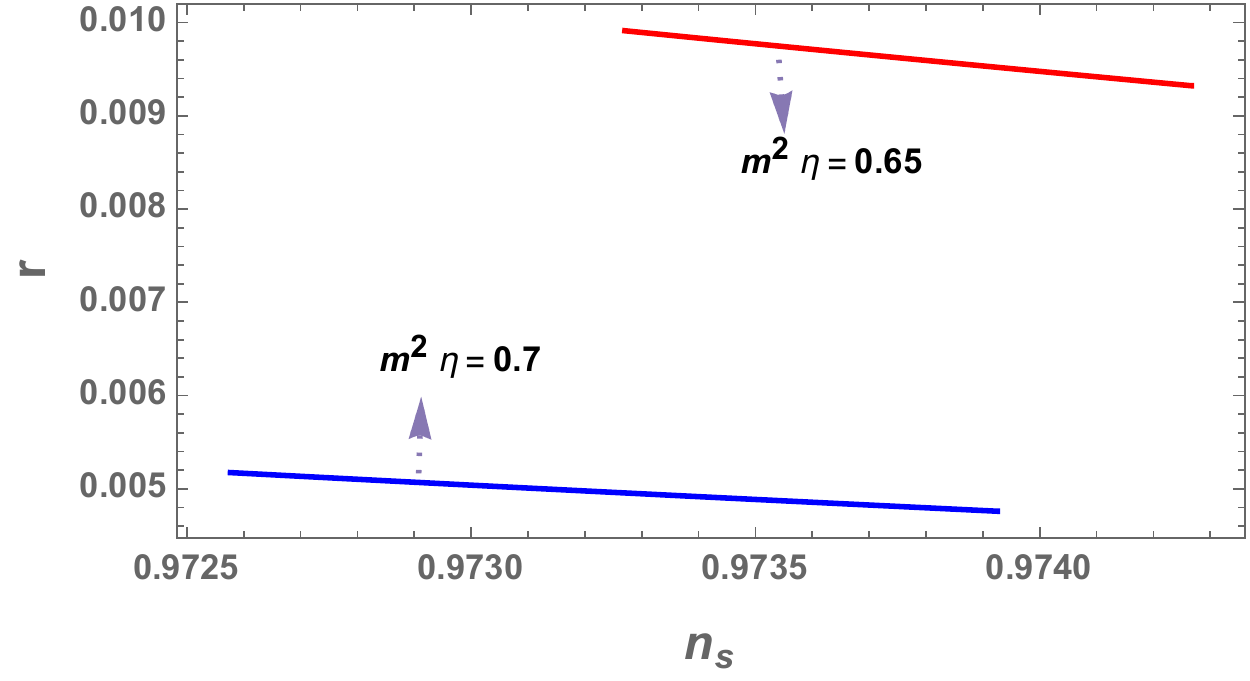}
\caption{The behavior of the scalar spectral index $n_s$ and $r$ for $m^2\gamma$, in the interval $1\le m^2\gamma\le 5$ and  $m^2\eta=0,65$ , $m^2\eta=0.7$ (in units of $M_p^4$).}
\end{figure}


\noindent {\bf Model III.}

\noindent The following model considers the general power-law potential and non-minimal power-law functions for the GB and kinetic couplings 
\be\label{ejec1}
F=\frac{1}{\kappa^2}, \;\;\; V(\phi)=\frac{\lambda}{n}\phi^n,\;\;\; F_1(\phi)=\frac{\gamma}{\phi^n}, \;\;\; F_2(\phi)=\frac{\eta}{\phi^n}
\ee
The slow-roll parameters (\ref{eqm8})-(\ref{eqm11}) for this model take the form ($\kappa=1$)
\be\nonumber
\epsilon_0=\frac{n^2(3n-8\eta\lambda)}{6(n+2\gamma\lambda)\phi^2},\;\;\; \epsilon_1=\frac{2n(3n-8\eta\lambda)}{3(n+2\gamma\lambda)\phi^2},\;\;\; 
\Delta_0=\frac{8n\eta\lambda(3n-8\eta\lambda)}{9(n+2\gamma\lambda)\phi^2},
\ee
\be\label{ejec2}
\Delta_1=\frac{2n(3n-8\eta\lambda)}{3(n+2\gamma\lambda)\phi^2},\;\;\; k_0=\frac{n\gamma\lambda(3n-8\eta\lambda)^2}{9(n+2\gamma\lambda)\phi^2},\;\;\; k_1=\frac{2n(3n-8\eta\lambda)}{3(n+2\gamma\lambda)\phi^2}
\ee
The scalar field at the end of inflation ($\epsilon_0=1$) takes the form
\be\label{ejec3}
\phi_E=\frac{n\sqrt{3n-8\eta\lambda}}{\sqrt{6n+12\gamma\lambda}}
\ee
The number of e-foldings from (\ref{eqm17}) is given by
\be\label{ejec4}
 N=-\frac{3(n+2\gamma\lambda)}{2n(3n-8\eta\lambda)}\phi^2\Big|_{\phi_I}^{\phi_E}
 \ee
which, using (\ref{ejec3}) allows to find the exact explicit form for the scalar field $N$ $e$-folds before the end of inflation as
\be\label{ejec5}
\phi_I=\left(\frac{(4N+n)(3n^2-8n\eta\lambda)}{6n+12\gamma\lambda}\right)^{1/2}=\sqrt{\frac{\left(4N+n\right)}{n}}\phi_E
\ee
From (\ref{slr29}) and (\ref{ejec2}) after replacing the value of the scalar field $\phi_I$ from (\ref{ejec5}), we find the scalar spectral index as
\be\label{ejec6}
n_s=\frac{\left(6\gamma\lambda+16n\eta\lambda+3n\right)\phi^2-3n^3-(6-8\eta\lambda)n^2}{3(n+2\gamma\lambda)\phi^2} \Big|_{\phi_I}=\frac{4N-n-4}{4N+n}
\ee
And from (\ref{slrt16}), (\ref{ejec2}) and (\ref{ejec5}) we find the expression for the tensor-to-scalar ratio as
\be\label{ejec7}
r=\frac{8n(3n-8\eta\lambda)^2}{9(n+2\gamma\lambda)\phi^2}\Big|_{\phi_I}=\frac{16(3n-8\eta\lambda)}{3(4N+n)}
\ee
The slow-roll parameters $N$ $e$-folds before the end of inflation take the values
\be\nonumber
\epsilon_0=\frac{n}{4N+n},\;\;\; \epsilon_1=\frac{4}{4N+n},\;\;\; \Delta_0=\frac{16\eta\lambda}{3(4N+n)},
\ee
\be\label{ejec8}
\Delta_1=\frac{4}{4N+n},\;\;\; k_0=\frac{2\gamma\lambda(3n-8\eta\lambda)}{3(4N+n)(n+2\gamma\lambda)},\;\;\; k_1=\frac{4}{4N+n}
\ee
The Eq. (\ref{ejec6}) predicts the scalar spectral index $n_s$ in terms of the number of $e$-foldings N, and the power $n$, which is the same result as that obtained for the standard chaotic inflation. However, the tensor-to-scalar ratio depends additionally on the self coupling $\lambda$ and the GB coupling constant $\eta$, but not on the kinetic coupling constant. As can be seen from the expressions (\ref{ejec3}) and  (\ref{ejec5}), the kinetic coupling can lower the values of the scalar field at the end, and therefore at the beginning, of inflation. Note also that the strong coupling regime of the GB coupling spoils the inflation ($\Delta_0$ and $k_0$ break the slow-roll restrictions), while at the strong coupling limit all slow-roll parameters and derived quantities are well defined.
Note also that all of the slow roll parameters (\ref{ejec2}), and therefore the quantities derived from them, depend on coupling constants through the products $\eta\lambda$ and $\gamma\lambda$. The dimension of $\eta$ is $mass^{n}$, the dimension of $\lambda$ is $mass^{4-n}$ and the dimension of $\gamma$ is $mass^{n-2}$, and therefore independently of $n$, the  product $\eta\lambda$ has constant dimension $[\eta\lambda]=mass^4$ and the corresponding dimension of $\gamma\lambda$ is $mass^2$ . This can be used to write $\eta\lambda=\alpha M_p^4$ where $\alpha$ is a dimensionless parameter that defines the behavior of $r$ once $n$ and $N$ have been fixed. While the coupling $\lambda$ is subject to different restrictions, depending on the power $n$, one can vary the coupling $\eta$ (and therefore $\alpha$) to find the appropriate value for the tensor-to-scalar ratio. On the other hand, the parameter $\beta=\gamma\lambda$ leads to consistent inflation in the weak coupling, $\gamma\rightarrow 0$, and strong coupling, $\gamma\rightarrow\infty$, limits and can take any value between these limits. 
In table I we list some sample values for $n_s$, $r$, for $N=60$ and a range of $\alpha$, for some power-law models including models with fractional $n$ that appear in string theory compactification \cite{silverstein, mcallister} and are favored by Planck 2018 data \cite{planck18}.\\

\begin{center}
\begin{tabular}{lccc}\hline\hline
Power $n$      & $n_s$       &Parameter $\alpha$ range       & $r$ in $\alpha$ range   \\
\hline
4 & 0.9508 & $1\le\alpha\le 1.3$& $0.0874\ge r\ge 0.0349$\\

3&0.959& $0.7\le\alpha\le1.1$&$ 0.0746\ge r\ge 0.0044$\\

2&0.9669& $0.3\le\alpha\le0.7$&$ 0.0746\ge r\ge 0.0088$\\
4/3& 0.9724&$0.01\le\alpha\le 0.4$&$ 0.0866\ge r\ge 0.0177$\\
1& 0.9751&$10^{-3}\le\alpha\le 0.3$&$ 0.0664\ge r\ge 0.0133$\\
2/3& 0.9778&$10^{-4}\le\alpha\le 0.2$&$ 0.0443\ge r\ge 0.0089$\\
\hline
\end{tabular}
\end{center}



\noindent {\bf Table I}. Some values of $n_s$ and $r$ in an appropriate range for $\alpha$ in each case.\\

\noindent It is noticeable the $n=2$ case, which for minimally coupled scalar field is disfavored by the latest observations \cite{planck18, bkp}, but in the presence of GB coupling falls in the range favored by the observational data. For all cases, the low tensor-to-scalar ratio is consistent with current observations. Since the parameter $\beta=\gamma\lambda$ is a free parameter, then one can use this freedom to set the values $\phi_E$, and therefore $\phi_I$, to any desired value.\\

\noindent {\bf Model IV.}
\noindent This model considers the general power-law potential and the non-minimal kinetic coupling of the form
\be\label{ejemd1}
F=\frac{1}{\kappa^2}, \;\;\; V(\phi)=\frac{\lambda}{n}\phi^n,\;\;\; F_1(\phi)=\frac{\beta}{\phi^{n+2}}, \;\;\; F_2(\phi)=0
\ee
The slow-roll parameters (\ref{eqm8})-(\ref{eqm11}) take the form ($\kappa=1$)
\be\nonumber
\epsilon_0=\frac{n^3}{2n\phi^2+4\beta\lambda\phi^4},\;\;\; \epsilon_1=\frac{2n^2(n+4\beta\lambda\phi^2)}{\phi^2(n+2\beta\lambda\phi^2)^2}
\ee
\be\label{ejemd2}
k_0=\frac{\beta\lambda n^3}{(n+2\beta\lambda\phi^2)^2},\;\;\; k_1=\frac{8\beta\lambda n^2}{(n+2\beta\lambda\phi^2)^2}.
\ee
from above the first equation we fond the scalar field at the end of inflation as
\be\label{ejemd3}
\phi_E^2=\frac{\sqrt{n^2(1+4n\beta\lambda)}-n}{4\beta\lambda}.
\ee
The number of e-foldings from (\ref{eqm17}) is
\be\label{ejemd4}
N=\frac{\beta\lambda\phi^4}{2n^2}-\frac{\phi^2}{2n}\Big|_{\phi_I}^{\phi_E}
\ee
This equation allows to find the scalar field $N$ $e$-foldings before the end of inflation as
\be\label{ejemd5}
\phi_I^2=\frac{n}{4\beta\lambda}\left(\sqrt{2}n\sqrt{\frac{n+2n\beta\lambda( n+8N)+n\sqrt{1+4n\beta\lambda}}{n^3}}-2\right)
\ee
Using this result we find the expression for the scalar spectral index from (\ref{slr29}) and (\ref{ejemd2}) as
\be\label{ejemd6}
n_s=1-\frac{4n\beta\lambda\left[\sqrt{2}n(n+4) f(n,N,\beta,\lambda)-4\right]}{n^3f^2(n,N,\beta,\lambda)\left[\sqrt{2}n f(n,N,\beta,\lambda)-2\right]},
\ee
where
$$ f(n,N,\beta,\lambda)=\sqrt{\frac{n+2n\beta\lambda( n+8N)+n\sqrt{1+4n\beta\lambda}}{n^3}}.$$
An for the tensor-to-scalar ratio it is found (from (\ref{slrt16}), (\ref{ejemd2}) and (\ref{ejemd5}))
\be\label{ejemd7}
r=\frac{32\sqrt{2}\beta\lambda}{f(n,N,\beta,\lambda)\left[\sqrt{2} n f(n,N,\beta,\lambda)-2\right]}
\ee
As can be seen form above results, both the slow-roll parameters and all the observable quantities depend on the product $\beta\lambda$, which independently of the power $n$, has dimensions of $(mass)^4$. The coupling $\lambda$ takes different significance and undergoes different restrictions depending on $n$, but we have some freedom in choosing the coupling $\beta$, so we can define the free parameter
\be\label{ejemd8}
\alpha=\beta\lambda
\ee
In table II we list some  sample values for the power-law potentials considered in Table I. $N=60$ is assumed and an appropriate range of $\alpha$ is chosen for each power $n$.\\

\begin{center}
\begin{tabular}{lccc}\hline\hline
Power $n$      & $n_s$       &Parameter $\alpha$ range       & $r$ in $\alpha$ range   \\
\hline
4 & $0.9666\le n_s\le 0.9667$ & $10\le\alpha\le 20$& $0.1335\ge r\ge 0.1339$\\

3& $0.9709\le n_s\le 0.971$ & $10^2\le\alpha\le 10^3$&$ 0.0998\ge r\ge 0.0995$\\

2& $0.973\le n_s\le 0.9744$ & $0.1\le\alpha\le 1$&$ 0.0756\ge r\ge 0.0692$\\

4/3& $0.9721\le n_s\le 0.9736$ &$10^{-3}\le\alpha\le 10^{-2}$&$ 0.080\ge r\ge 0.062$\\

1& $0.9746\le n_s\le 0.977$ &$10^{-3}\le\alpha\le 0.05$&$ 0.06\ge r\ge 0.04$\\

2/3& $0.9777\ge n_s\ge 0.9774 $&$10^{-4}\le\alpha\le 10^{-2}$&$ 0.0438\ge r\ge 0.0313$\\
\hline
\end{tabular}
\end{center}
\noindent {\bf Table II}. $n_s$ and $r$ in an appropriate range for $\alpha$ in each case.\\

\noindent Notice that $n_s$ varies in very narrow intervals, retaining almost the same value in each case. The quartic potential presents better values for $n_s$ compared to the previous model, but the tensor-to-scalar ratio becomes larger that in the previous model, moving away from the values favored by the latest observations. The quadratic potential maintains its viability in the present model, although $r$ increases a bit with respect to the model (\ref{ejec1}).
From the expressions (\ref{ejemd6}) and (\ref{ejemd7}) we find the following behavior for $n_s$ and $r$ in the strong coupling limit ($\beta\rightarrow\infty$) 
\be\label{ejemd9}
\lim_{\beta\to\infty} n_s=\frac{8N-n-8}{8N+n},\;\;\; \lim_{\beta\to\infty}r=\frac{16n}{8N+n}
\ee
In the weak coupling limit, $\beta\rightarrow 0$, it is found
\be\label{ejemd10}
\lim_{\beta\to 0} n_s=\frac{4N-n-4}{4N+n},\;\;\; \lim_{\beta\to 0}r=\frac{16n}{4N+n}.
\ee
From the expressions for $\phi_E$ and $\phi_I$ we find that at the strong coupling limit
\be\label{ejemd12}
\phi_E\rightarrow \left(\frac{n^3}{4\beta\lambda}\right)^{1/4},\;\;\; \phi_I\rightarrow \left(\frac{n^2(8N+n)}{4\beta\lambda}\right)^{1/4}
\ee
and at the weak coupling limit, from the slow-roll parameter $\epsilon_0$ and $N$ from (\ref{ejemd4}), the $\phi_E$ and $\phi_I$ fields tend to the constant values
\be\label{ejemd13}
\phi_E\rightarrow \frac{n}{\sqrt{2}},\;\;\; \phi_I\rightarrow \frac{n\sqrt{4N+n}}{\sqrt{2}}
\ee
It is clear that in the strong coupling regime the scalar field at the beginning and end of inflation takes smaller values compared to the standard chaotic inflation.
\section{Discussion}
The slow-roll inflation driven by a single scalar field with non-minimal couplings of different nature, that lead to second order field equations, have been studied. The detailed analysis of the linear and quadratic perturbations for all the interaction terms in the model is given. The second oder action for scalar and tensor perturbations have been constructed, and the expressions for the scalar and tensor power spectra in terms of the slow-roll parameters have been obtained. In Eq. (\ref{slrt18}) we give the consistency relation that allows to discriminate the model from the standard inflation with minimally coupled scalar field. The results were applied to some models with power-law potential. For the scalar field with quadratic potential, non-minimal coupling and kinetic coupling to the Einstein tensor (\ref{ejea1}), we have found that the scalar spectral index and the tensor-to-scalar ratio can take values in the region favored by the latest observational data \cite{planck18, bkp}, as seen in Fig. 2.
The quadratic potential is also considered with kinetic and GB couplings (\ref{ejeb1}). In this case $n_s$ takes values in the region $0.972\lesssim n_s\lesssim0.974$, and the range of values for $r$ is of the order $\approx 10^{-2}$ for $60$ $e$-foldings, which falls in the region bounded by \cite{planck18, bkp}. A general monomial potential $V\propto \phi^n$ with non-minimal kinetic coupling, $F_1\propto\phi^{-n}$ and non-minimal GB coupling $F_2\propto \phi^{-n}$, was considered (\ref{ejec1}). For this model it was possible to find exact analytical expressions for the main quantities in the slow-roll approximation, and some notable values of $n$ were analyzed. While the predictions for $n_s$ correspond to the standard chaotic inflation, the results for $r$ could be improved due to the GB coupling, and particularly, for the quadratic potential it was found that the tensor-to-scalar ratio falls in an appropriate range according to the latest restrictions, as can be seen in table I. 
Analyzing the behavior of the model (\ref{ejec1}) in the weak and strong coupling limits, it was shown that the inflation is not viable in the strong GB coupling limit, especially because $\Delta_0$ and $k_0$ break the slow-roll restrictions (see (\ref{ejec8})) and the tensor-to-scalar ratio (\ref{ejec7}) increases substantially, while the kinetic coupling remains consistent with inflation in the strong coupling limit. The kinetic coupling constant, as a free parameter, can be used to lower the value of the scalar field to any desired value at the end, and therefore at the beginning, of inflation, avoiding in this way the problem of large fields in chaotic inflation.\\
Another interesting situation is found when we consider the model (\ref{ejemd1}) with a power-law potential $V\propto\phi^n$ and non-minimal kinetic coupling ($F_1=\beta/\phi^{n+2}$). In this case both $n_s$ and $r$ depend on the kinetic coupling constant and the model behaves appropriately for any value of the coupling between the weak and strong coupling regimes. In the weak coupling limit we recover the standard chaotic inflation results, and in the strong coupling limit we can see from (\ref{ejemd9}) and (\ref{ejemd10}) that $n_s$ increments with respect to its value in the weak limit, and $r$ decreases with respect to its value in the weak coupling limit. This effect is appreciable, in fact, in the intermediate regime as seen in table II. Thus, for the quadratic potential the tensor-to-scalar ratio falls in the region favored by the latest observations \cite{planck18, bkp}, since $n_s$ can reach a maximum value of $(4N-5)/(4N+1)$ and $r$ can reach the minimum value of $16/(4N+1))$. For the quartic potential $V=\lambda\phi^4/4$, $n_s$ can reach a maximum value of $(2N-3)/(2N+1)$ and $r$ reaches a minimum value of $16/(2N+1)$ which, assuming $N=60$ gives $r=0.1322$, which is lower than in the standard chaotic inflation, but is not enough to satisfy the restriction $r<0.1$. \\
The latest observational data disfavor monomial-type models $V\propto \phi^n$ with $n\ge 2$ in the minimally coupled scalar field. With the Introduction of additional interactions like the non-minimal coupling, kinetic coupling and Gauss-Bonnet coupling, it is shown that the tensor-to-scalar ratio can be lowered to values that are consistent with latest observational constraints \cite{planck18, bkp}. 
An important consequence of the kinetic coupling in (\ref{ejec1}) and (\ref{ejemd1}), is that the coupling parameter can take any value between the weak and strong coupling limits which gives rise to the freedom to impose any physical bounds on the self-coupling $\lambda$, depending on the power $n$.\\
It is clear that the inclusion of non-minimal kinetic and GB couplings in single scalar field inflationary scenarios has important consequences for the observable magnitudes, as was shown in the case of monomial potentials (see also \cite{ohashi, stsujikawa}, \cite{Kanti:1998jd}-\cite{vandeBruck:2017voa}). Further analysis of different single scalar field cosmological scenarios will be considered in the presence of these couplings.

\appendix

\section{Basic formulas for the first order perturbations}
To analyze the physical phenomena during the period of inflation and make contact with the observables that originated at that period, we start with the perturbations around the homogeneous FRW background of the scalar field and the metric (including the geometrical quantities derived from it) involved in the inflation. The metric with its first order perturbation is written as
\be\label{eqa1}
g_{\mu\nu}=\bar{g}_{\mu\nu}(t)+h_{\mu\nu}(\vec{x},t)
\ee 
where $\bar{g}_{\mu\nu}$ is the background FRW metric with components
\be\label{eqa2}
\bar{g}_{00}=-1,\;\;\; \bar{g}_{i0}=\bar{g}_{0i}=0,\;\;\; \bar{g}_{ij}=a(t)^2\delta_{ij}
\ee
and $h_{\mu\nu}=h_{\nu\mu}$ is the small perturbation of the metric which satisfies the following first order relation
\be\label{eqa3}
h^{\mu\nu}=-\bar{g}^{\mu\rho}\bar{g}^{\nu\lambda}h_{\rho\lambda},
\ee 
that follows from the metric property $g_{\mu\nu}g^{\nu\rho}=\delta^{\rho}_{\mu}$. Writing in components we find
\be\label{eqa4}
h^{ij}=-a(t)^{-4}h_{}ij,\;\;\; h^{i0}=a(t)^{-2}h_{i0},\;\;\; h^{00}=-h_{00}.
\ee
The background Christoffel symbols are given by
\be\label{eqa5}
\bar{\Gamma}^i_{j0}=\bar{\Gamma}^i_{0j}=\frac{\dot{a}}{a}\delta_{ij},\;\;\; \bar{\Gamma}^0_{ij}=a\dot{a}\delta{ij},\;\;\; \bar{\Gamma}^i_{jk}=0
\ee
performing the first order perturbation in the Christoffel symbols  for the metric (\ref{eqa1}) we find the following components 
\be\label{eqa6}
\delta\Gamma^i_{jk}=\frac{1}{2a^2}\left(-2a\dot{a}\delta_{jk}h_{i0}+\partial_k h_{ij}+\partial_j h_{ik}-\partial_i h_{jk}\right)
\ee
\be\label{eqa7}
\delta\Gamma^i_{j0}=\frac{1}{2a^2}\left(-2\frac{\dot{a}}{a}h_{ij}+\dot{h}_{ij}+\partial_j h_{i0}-\partial_i h_{j0}\right)
\ee
\be\label{eqa8}
\delta\Gamma^0_{ij}=\frac{1}{2}\left(2a\dot{a}h_{00}\delta_{ij}-\partial_i h_{0j}-\partial_j h_{0i}+\dot{h}_{ij}\right)
\ee
\be\label{eqa9}
\delta\Gamma^i_{00}=\frac{1}{2a^2}\left(2\dot{h}_{i0}-\partial_i h_{00}\right)
\ee
\be\label{eqa10}
\delta\Gamma^0_{i0}=\frac{\dot{a}}{a}h_{i0}-\frac{1}{2}\partial_i h_{00}
\ee
\be\label{eqa11}
\delta\Gamma^0_{00}=-\frac{1}{2}\dot{h}_{00}
\ee
and there is a useful formula for the trace of $\delta\Gamma$
\be\label{eqa12}
\delta\Gamma^{\lambda}_{\lambda\mu}=\partial_{\mu}\left(\frac{1}{2a^2}h_{ii}-\frac{1}{2}h_{00}\right)
\ee
In what follows all the calculations will be performed in the Newtonian gauge. The first-order perturbation formalism will be applied to all terms  
in the general scalar-tensor model described bellow, and here we describe the result for the basic geometrical quantities.  
In the Newtonian gauge, after the standard scalar-vector-tensor decomposition of the metric perturbations (see \cite{weinberg}), it is obtained 
$$B=F=0,\;\;\; E=2\Phi,\;\;\; A=-2\Psi, $$
And the metric perturbations take the form
\be\label{eqb2}
\begin{aligned}
&h_{00}=-2\Phi,\;\;\; h_{i0}=h_{0i}=0,\;\;\; h_{ij}=-2a^2\Psi\delta_{ij}\\&
 h^{00}=2\Phi,\;\;\; h^{i0}=h^{0i}=0,\;\;\; h^{ij}=2a^{-2}\Psi \delta_{ij}
\end{aligned}
\ee
Replacing these expressions into the results for the perturbations of the Christoffel symbols given in Eqs. (\ref{eqa6})-(\ref{eqa12}) we find
\be\label{eqb3}
\begin{aligned}
&\delta\Gamma^0_{00}=\dot{\Phi},\;\;\; \delta\Gamma^0_{i0}=-\partial_i\Phi,\;\;\; \delta\Gamma^i_{00}=\frac{1}{a^2}\partial_i\Phi,\;\;\;
\delta\Gamma^0_{ij}=-2a\dot{a}\Phi\delta_{ij}-2a\dot{a}\Psi\delta_{ij}-a^2\dot{\Psi}\delta{ij}\\& \delta\Gamma^i_{j0}=-\dot{\Psi}\delta_{ij},\;\;\;
\delta\Gamma^i_{jk}=-\partial_k\Psi\delta_{ij}-\partial_j\Psi\delta_{ik}+\partial_i\Psi\delta_{jk}
\end{aligned}
\ee
For the curvature tensor
\be\label{eqb4}
R^{\rho}_{\sigma\mu\nu}=\partial_{\mu}\Gamma^{\rho}_{\sigma\nu}-\partial_{\nu}\Gamma^{\rho}_{\sigma\mu}+\Gamma^{\rho}_{\mu\lambda}\Gamma^{\lambda}_{\sigma\nu}-\Gamma^{\rho}_{\nu\lambda}\Gamma^{\lambda}_{\sigma\mu},
\ee
The background components are given by
\be\label{eqb5}
\begin{aligned}
&\bar{R}^i_{0j0}=-\bar{R}^i_{00j}=-\left(H^2+\dot{H}\right),\;\;\; \bar{R}^0_{i0j}=-\bar{R}^0_{ij0}=a\ddot{a}\delta_{ij},\;\;\;  \bar{R}^i_{jk0}=0
\\&\bar{R}^i_{jkl}=\dot{a}^2\left(\delta_{ik}\delta_{lj}-\delta_{il}\delta_{jk}\right),\;\;\;
\bar{R}^0_{000}=\bar{R}^0_{i00}=\bar{R}^0_{0i0}=\bar{R}^0_{00i}=\bar{R}^i_{000}=0.
\end{aligned}
\ee
The first order perturbations are given by
\be\label{eqb6}
\delta R^{\rho}_{\sigma\mu\nu}=\partial_{\mu}\delta\Gamma^{\rho}_{\sigma\nu}-\partial_{\nu}\delta\Gamma^{\rho}_{\sigma\mu}+\bar{\Gamma}^{\rho}_{\mu\lambda}\delta\Gamma^{\lambda}_{\sigma\nu}+\delta\Gamma^{\rho}_{\mu\lambda}\bar{\Gamma}^{\lambda}_{\sigma\nu}-\bar{\Gamma}^{\rho}_{\nu\lambda}\delta\Gamma^{\lambda}_{\sigma\mu}--\delta\Gamma^{\rho}_{\nu\lambda}\bar{\Gamma}^{\lambda}_{\sigma\mu}
\ee
Using (\ref{eqa5}) and (\ref{eqb3}) in (\ref{eqb6}) we find the first-order perturbations for the components of the curvature tensor
\be\label{eqb7}
\delta R^i_{0j0}=-\delta R^i_{00j}=\frac{1}{a^2}\partial_i\partial_j\Phi+\left(\ddot\Psi+H\dot{\Phi}+2H\dot{\Psi}\right)\delta_{ij}
\ee
\be\label{eqb8}
\delta R^i_{jk0}=-\delta R^i_{j0k}=\partial_j\dot{\Psi}\delta_{ik}-\partial_i\dot{\Psi}\delta_{jk}+H\partial_j\Phi\delta_{ik}-H\partial_i\Phi\delta_{jk}
\ee
\be\label{eqb9}
\delta R^0_{i0j}=-\delta R^0_{ij0}=-\left(2a\ddot{a}\Phi+a\dot{a}\dot{\Phi}+2a\ddot{a}\Psi+a^2\ddot{\Psi}\right)\delta_{ij}-\partial_i\partial_j\Phi
\ee
\be\label{eqb10}
\begin{aligned}
\delta R^i_{jkl}&=-\partial_k\partial_j\Psi\delta_{il}+\partial_k\partial_i\Psi\delta_{jl}+\partial_l\partial_j\Psi\delta_{ik}-\partial_l\partial_i\Psi\delta_{jk}-2a\dot{a}\dot{\Psi}\delta_{ik}\delta_{lj}\\&-2\dot{a}^2\Phi\delta_{ik}\delta_{lj}
-2\dot{a}^2\Psi\delta_{ik}\delta_{lj}+2a\dot{a}\dot{\Psi}\delta_{il}\delta_{kj}+2\dot{a}^2\Phi\delta_{il}\delta_{kj}+2\dot{a}^2\Psi\delta_{il}\delta_{kj}
\end{aligned}
\ee
\be\label{eqb11}
\delta R^0_{000}=\delta R^0_{i00}=\delta R^0_{0i0}=\delta R^i_{000}=\delta R^0_{00i}=0
\ee
Contracting (\ref{eqb6}) we find the different components of the perturbation of Ricci tensor as
\be\label{eqb12}
\delta R_{00}=\frac{1}{a^2}\nabla^2\Phi+3\ddot{\Psi}+3H\dot{\Phi}+6\dot{H}\dot{\Psi},
\ee
\be\label{eqb13}
\delta R_{i0}=2\partial_i\dot{\Psi}+2H\partial_i\Phi,
\ee
\be\label{eqb14}
\begin{aligned}
\delta R_{ij}=&-\left(2a\ddot{a}\Phi+4\dot{a}^2\Phi+a\dot{a}\dot{\Phi}+4\dot{a}^2\Psi+2a\ddot{a}\Psi+6a\dot{a}\dot{\Psi}+a^2\ddot{\Psi}\right)\delta_{ij}\\& -\partial_i\partial_j\Phi+\partial_i\partial_j\Psi+\nabla^2\Psi\delta_{ij}.
\end{aligned}
\ee
For the mixed components it is found
\be\label{eqb15}
\delta R^{0}_{0}=-6H^2\Phi-6\dot{H}\Phi-\frac{1}{a^2}\nabla^2\Phi-3H\dot{\Phi}-6H\dot{\Psi}-3\ddot{\Psi}
\ee
\be\label{eqb16}
\delta R^{0}_i=-2\partial_i\left(H\Phi+\dot{\Psi}\right)
\ee
\be\label{eqb17}
\delta R^{i}_j=-\left(6H^2\Phi+2\dot{H}\Phi+H\dot{\Phi}+6H\dot{\Psi}+\ddot{\Psi}\right)\delta_{ij}+\frac{1}{a^2}\nabla^2\Psi\delta_{ij}-\frac{1}{a^2}\partial_i\partial_j\left(\Phi-\Psi\right)
\ee
And the perturbation for the scalar curvature is given by  
\be\label{eqb18}
\delta R=-12\left(2H^2+\dot{H}\right)\Phi-6H\dot{\Phi}-\frac{2}{a^2}\nabla^2\Phi-24H\dot{\Psi}-6\ddot{\Psi}+\frac{4}{a^2}\nabla^2\Psi
\ee
For the scalar-tensor models that involve non-minimal couplings of the scalar field to curvatures, given by general functions $f(\phi)$, the energy momentum tensor contains covariant derivatives of these functions of the scalar field. Here we give the perturbations for expressions that involve two covariant derivatives of functions of the scalar field. Let's consider the following derivatives
\be\label{eqb19}
\nabla_{\mu}\nabla_{\nu}f(\phi)=\partial_{\mu}\partial_{\nu}f(\phi)-\Gamma^{\lambda}_{\mu\nu}\partial_{\lambda}f(\phi),\;\;\; \nabla^{\mu}\nabla_{\nu}f(\phi)=g^{\mu\lambda}\nabla_{\lambda}\nabla_{\nu}f(\phi).
\ee
Then
\be\label{eqb20}
\begin{aligned}
&\nabla_0\nabla_0f(\phi)=\partial_{0}\partial_{0}f(\phi)=\dot{\phi}^2f''(\phi)+\ddot{\phi}f'(\phi)\\&
\nabla_0\nabla_if(\phi)=\nabla_i\nabla_0f(\phi)=0\\&
\nabla_i\nabla_jf(\phi)=-a\dot{a}\dot{\phi}f'(\phi)\delta_{ij}\\&
 \nabla^{0}\nabla_{0}f(\phi)=-\dot{\phi}^2f''(\phi)-\ddot{\phi}f'(\phi)\\&
 \nabla^0\nabla_if(\phi)=-\nabla_0\nabla_if(\phi)=0\\&
\nabla^i\nabla_jf(\phi)=-H\dot{\phi}f'(\phi)\delta_{ij}\\&
 \nabla^{\mu}\nabla_{\mu}f(\phi)=-3H\dot{\phi}f'(\phi)-\dot{\phi}^2f''(\phi)-\ddot{\phi}f'(\phi)\\&
\nabla^0\nabla^0f(\phi)=\dot{\phi}^2f''(\phi)+\ddot{\phi}f'(\phi)\\& 
\nabla^0\nabla^i f(\phi)=\nabla^i\nabla^0 f(\phi)=0 \\&
\nabla^i\nabla^j f(\phi)=-\frac{\dot{a}}{a^3}\dot{\phi}f'(\phi)\delta_{ij}.
\end{aligned}
\ee
Here ' represents derivative w.r.t. the scalar field $\phi$. Let us consider the perturbations of the above derivative terms
\be\label{eqb21}
\delta\Big[\nabla_{\mu}\nabla_{\nu}f(\phi)\Big]=\partial_{\mu}\partial_{\nu}\Big[f'(\phi)\delta\phi\Big]-\delta\Gamma^{\lambda}_{\mu\nu}\partial_{\lambda}f(\phi)-\bar{\Gamma}^{\lambda}_{\mu\nu}\partial_{\lambda}\Big[f'(\phi)\delta\phi\Big] 
\ee
For the different components we find
\be\label{eqb22}
\delta\left[\nabla_{0}\nabla_{0}f(\phi)\right]=\dot{\phi}^2f'''(\phi)\delta\phi+\ddot{\phi}f''(\phi)\delta\phi+2\dot{\phi}f''(\phi)\dot{\delta\phi}-\dot{\Phi}f'(\phi)\dot{\phi}
\ee
\be\label{eqb23}
\delta\Big[\nabla_0\nabla_if(\phi)\Big]=f''(\phi)\dot{\phi}\partial_i\delta\phi+f'(0)\partial_0\partial_i\delta\phi-\frac{\dot{a}}{a}f'(\phi)\partial_i\delta\phi-f'(\phi)\dot{\phi}\partial_i\Phi
\ee
\be\label{eqb24}
\begin{aligned}
\delta\Big[\nabla_i\nabla_jf(\phi)\Big]=&f'(\phi)\partial_{i}\partial_{j}\delta\phi+\left(2a\dot{a}\Phi+2a\dot{a}\Psi+a^2\dot{\Psi}\right)\dot{\phi}f'(\phi)\delta_{ij}\\&-a\dot{a}\left(f''(\phi)\dot{\phi}\delta\phi+f'(\phi)\dot{\delta\phi}\right)\delta{ij}
\end{aligned}
\ee
\be\label{eqb25}
\begin{aligned}
\delta\Big[ \nabla^{0}\nabla_{0}f(\phi)\Big]=&-\left(f'''(\phi)\dot{\phi}^2\delta\phi+f''(\phi)\ddot{\phi}\delta\phi+2f''(\phi)\dot{\phi}\dot{\delta\phi}+f'(\phi)\ddot{\delta\phi}-f'(\phi)\dot{\phi}\dot{\Phi}\right)\\&+2\left(f''(\phi)\dot{\phi}^2+f'(\phi)\ddot{\phi}\right)\Phi
\end{aligned}
\ee
\be\label{eqb26}
\delta \Big[\nabla^0\nabla_if(\phi)\Big]=-f''(\phi)\dot{\phi}\partial_i\delta\phi-f'(\phi)\partial_i\dot{\delta\phi}+Hf'(\phi)\partial_i\delta\phi+f'(\phi)\dot{\phi}\partial_i\Phi
 \ee
\be\label{eqb27}
\begin{aligned}
\delta\Big[\nabla^i\nabla_jf(\phi)\Big]=&\frac{1}{a^2}f'(\phi)\partial_i\partial_j\delta\phi+\\&\left(2Hf'(\phi)\dot{\phi}\Phi+f'(\phi)\dot{\phi}\dot{\Psi}-Hf''(\phi)\dot{\phi}\delta\phi-Hf'(\phi)\dot{\delta\phi}\right)\delta_{ij}
\end{aligned}
\ee
\be\label{eqb28}
\begin{aligned}
\delta\Big[ \nabla^{\mu}\nabla_{\mu}f(\phi)\Big]&=-f'''(\phi)\dot{\phi}^2\delta\phi-f''(\phi)\ddot{\phi}\delta\phi-2f''(\phi)\dot{\phi}\dot{\delta\phi}-f'(\phi)\ddot{\delta\phi}+f'(\phi)\dot{\phi}\dot{\Phi}+\\&2f''(\phi)\dot{\phi}^2\Phi+
2f'(\phi)\ddot{\phi}\Phi+\frac{f'(\phi)}{a^2}\nabla^2\delta\phi+6Hf'(\phi)\dot{\phi}\Phi+3f'(\phi)\dot{\phi}\dot{\Psi}-\\&
3Hf''(\phi)\dot{\phi}\delta\phi-3Hf'(\phi)\dot{\delta\phi}
\end{aligned}
\ee
\be\label{eqb29}
\begin{aligned}
\delta\Big[ \nabla^{0}\nabla^{0}f(\phi)\Big]&=-4\Phi\left(f''(\phi)\dot{\phi}^2+f'(\phi)\ddot{\phi}\right)+f'''(\phi)\dot{\phi}^2\delta\phi+f''(\phi)\ddot{\phi}\delta\phi \\&+2f''(\phi)\dot{\phi}\dot{\delta\phi}+f'(\phi)\ddot{\delta\phi}-f'(\phi)\dot{\phi}\dot{\Phi}
\end{aligned}
\ee
\be\label{eqb30}
\delta \Big[\nabla^0\nabla^if(\phi)\Big]=\frac{1}{a^2}\left(-f''(\phi)\dot{\phi}\partial_i\delta\phi-f'(\phi)\partial_i\dot{\delta\phi}+Hf'(\phi)\partial_i\delta\phi+f'(\phi)\dot{\phi}\partial_i\Phi\right)
\ee
\be\label{eqb31}
\begin{aligned}
\delta\Big[\nabla^i\nabla^j f(\phi)\Big]&=-\frac{2}{a^2}H f'(\phi)\dot{\phi}\Psi\delta_{ij}+\frac{1}{a^2}\Big(\frac{f'(\phi)}{a^2}\partial_i\partial_j\delta\phi+2H f'(\phi)\dot{\phi}\Phi\delta_{ij}
\\&+ f'(\phi)\dot{\phi}\dot{\Psi}\delta_{ij}-H f''(\phi)\dot{\phi}\delta\phi\delta_{ij}-H f'(\phi)\dot{\delta\phi}\delta_{ij}\Big)
\end{aligned}
\ee
\section{The scalar-tensor model and the equations of motion}

\noindent Using the above basic results for the fundamental geometrical quantities, we can proceed to evaluate the fist order perturbations 
for the following scalar-tensor model with non-minimal coupling to scalar curvature $R$, non-minimal kinetic coupling to the Ricci and scalar curvature through the Einstein tensor $G_{\mu\nu}$ and non-minimal coupling to the 4-dimensional Gauss-Bonnet invariant ${\cal G}$
\be\label{eq1c}
S=\int d^4x\sqrt{-g}\left[\frac{1}{2}F(\phi)R-\frac{1}{2}g^{\mu\rho}\partial_{\mu}\phi\partial_{\rho}\phi-V(\phi)+F_1(\phi)G_{\mu\nu}\partial^{\mu}\phi\partial^{\nu}\phi- F_2(\phi){\cal G}\right]
\ee
where $${\cal G}=R^2-4R_{\mu\nu}R^{\mu\nu}+R_{\mu\nu\lambda\rho}R^{\mu\nu\lambda\rho},$$ 
$$G_{\mu\nu}=R_{\mu\nu}-\frac{1}{2}g_{\mu\nu}R$$
and
$$F(\phi)=\frac{1}{\kappa^2}+f(\phi)$$.
To obtain the field equations we use the following basic variations 
\begin{equation}\label{var1}
\delta {g_{\mu \nu }} =  - {g_{\mu \rho }}{g_{\nu \sigma }}\delta {g^{\rho \sigma }},
\end{equation}
\begin{equation}\label{var2}
\delta \sqrt { - g}  =  - \frac{1}{2}\sqrt { - g} {g_{\mu \nu }}\delta {g^{\mu \nu }},
\end{equation}
\begin{equation}\label{var3}
\delta R = {R_{\mu \nu }}\delta {g^{\mu \nu }} + {g_{\mu \nu }}{\nabla _\sigma }{\nabla ^\sigma }\delta {g^{\mu \nu }} - {\nabla _\mu }{\nabla _\nu }\delta {g^{\mu \nu }},
\end{equation}
\begin{equation}\label{var4}
\delta {R_{\mu \nu }} = \frac{1}{2}\left( {{g_{\mu \alpha }}{g_{\nu \beta }}{\nabla _\lambda }{\nabla ^\lambda }\delta {g^{\alpha \beta }} + {g_{\alpha \beta }}{\nabla _\nu }{\nabla _\mu }\delta {g^{\alpha \beta }} - {g_{\mu \beta }}{\nabla _\alpha }{\nabla _\nu }\delta {g^{\alpha \beta }} - {g_{\nu \alpha }}{\nabla _\beta }{\nabla _\mu }\delta {g^{\alpha \beta }}} \right),
\end{equation}
\be\label{var5}
\begin{aligned}
\delta {R_{\alpha \beta \kappa \lambda }} = &\frac{1}{2}\Big({\nabla _\kappa }{\nabla _\beta }\delta {g_{\lambda \alpha }} + {\nabla _\lambda }{\nabla _\alpha }\delta {g_{\kappa \beta }} - {\nabla _\kappa }{\nabla _\alpha }\delta {g_{\lambda \beta }} - {\nabla _\lambda }{\nabla _\beta }\delta {g_{\kappa \alpha }}\nonumber\\
&  + R_{\beta \kappa \lambda }^\gamma \delta {g_{\gamma \alpha }} - R_{\alpha \kappa \lambda }^\gamma \delta {g_{\beta \gamma }}\Big).
\end{aligned}
\ee
The variation of the GB term requires, additionally, the use of the following Bianchi-related identities
\\
\begin{equation}\label{1}
{\nabla ^\rho }{R_{\rho \sigma \mu \nu }} = {\nabla _\mu }{R_{\sigma \nu }} - {\nabla _\nu }{R_{\sigma \mu }}
\end{equation}
\begin{equation}\label{2}
{\nabla ^\rho }{R_{\rho \mu }} = \frac{1}{2}{\nabla _\mu }R
\end{equation}
\begin{equation}\label{3}
{\nabla ^\rho }{\nabla ^\sigma }{R_{\sigma \rho }} = \frac{1}{2}\square R
\end{equation}
\begin{equation}\label{4}
{\nabla ^\rho }{\nabla ^\sigma }{R_{\mu \rho \nu \sigma }} = {\nabla ^\rho }{\nabla _\rho }{R_{\mu \nu }} - \frac{1}{2}{\nabla _\mu }{\nabla _\nu }R + {R_{\gamma \mu \lambda \nu }}{R^{\lambda \gamma }} - {R_{\gamma \mu }}R_\nu ^\gamma 
\end{equation}
\begin{equation}\label{5}
{\nabla ^\rho }{\nabla _\mu }{R_{\rho \nu }} + {\nabla ^\rho }{\nabla _\nu }{R_{\rho \mu }} = \frac{1}{2}\left( {{\nabla _\mu }{\nabla _\nu }R + {\nabla _\nu }{\nabla _\mu }R} \right) - 2{R_{\lambda \mu \gamma \nu }}{R^{\gamma \lambda }} + 2{R_{\lambda \nu }}R_\mu ^\lambda, 
\end{equation}
which can be obtained directly from the Bianchi identity.
\\
Variation with respect to metric gives the field equations
\be\label{eq2c}
R_{\mu\nu}-\frac{1}{2}g_{\mu\nu}R=\kappa^2 T_{\mu\nu}=\kappa^2\left( T^{\phi}_{\mu\nu}+ T^{NM}_{\mu\nu}+ T^{K}_{\mu\nu}+ T^{GB}_{\mu\nu}\right),
\ee
where
\be\label{eq3c}
\begin{aligned}
& T^{\phi}_{\mu\nu}=-\frac{2}{\sqrt{-g}}\frac{\delta S_{\phi}}{\delta g^{\mu\nu}},\;\;\; T^{NM}_{\mu\nu}=-\frac{2}{\sqrt{-g}}\frac{\delta S_{NM}}{\delta g^{\mu\nu}}\\ &
T^{K}_{\mu\nu}=-\frac{2}{\sqrt{-g}}\frac{\delta S_{K}}{\delta g^{\mu\nu}},\;\;\; T^{GB}_{\mu\nu}=-\frac{2}{\sqrt{-g}}\frac{\delta S_{GB}}{\delta g^{\mu\nu}},
\end{aligned}
\ee
with
\be\label{eq4c}
S_{\phi}=\int d^4x\sqrt{-g}\left[-\frac{1}{2}g^{\mu\rho}\partial_{\mu}\phi\partial_{\rho}\phi-V(\phi)\right],
\ee
\be\label{eq5c}
S_{NM}=\frac{1}{2}\int d^4x\sqrt{-g} f(\phi)R,
\ee
\be\label{eq6c}
S_{K}=\int d^4x\sqrt{-g}F_1(\phi)G_{\mu\nu}\partial^{\mu}\phi\partial^{\nu}\phi,
\ee
\be\label{eq7c}
S_{GB}=-\int d^4x\sqrt{-g}F_2(\phi){\cal G},
\ee
where
\be\label{eq8c}
T^{\phi}_{\mu\nu}=\partial_{\mu}\phi\partial_{\nu}\phi-\frac{1}{2}g_{\mu\nu}\partial_{\rho}\phi\partial^{\rho}\phi-g_{\mu\nu}V(\phi),
\ee
\be\label{eq10c}
T^{NM}_{\mu\nu}=-f(\phi)\left(R_{\mu\nu}-\frac{1}{2}g_{\mu\nu}R\right)-g_{\mu\nu}\nabla_{\sigma}\nabla^{\sigma}f(\phi)+\nabla_{\mu}\nabla_{\nu}f(\phi),
\ee
\be\label{eq11c}
\begin{aligned}
T_{\mu \nu }^K = &{F_1}{\partial _\rho }\phi {\partial ^\rho }\phi \left( {{R_{\mu \nu }} - \frac{1}{2}{g_{\mu \nu }}R} \right) + {g_{\mu \nu }}{\nabla ^\sigma }{\nabla _\sigma }\Big( {{F_1}{\partial _\rho }\phi {\partial ^\rho }\phi } \Big) - {\nabla _\nu }{\nabla _\mu }\Big( {{F_1}{\partial _\rho }\phi {\partial ^\rho }\phi } \Big)\\
& + {F_1}R{\partial _\mu }\phi {\partial _\nu }\phi  - 2{F_1}\Big( {{R_{\mu \rho }}{\partial _\nu }\phi {\partial ^\rho }\phi  + {R_{\nu \rho }}{\partial _\mu }\phi {\partial ^\rho }\phi } \Big) + {F_1}{g_{\mu \nu }}{R_{\rho \sigma }}{\partial ^\rho }\phi {\partial ^\sigma }\phi \\
& + {\nabla ^\rho }{\nabla _\mu }\Big( {{F_1}{\partial _\nu }\phi {\partial _\rho }\phi } \Big) + {\nabla ^\rho }{\nabla _\nu }\Big( {{F_1}{\partial _\mu }\phi {\partial _\rho }\phi } \Big) - {\nabla ^\sigma }{\nabla _\sigma }\Big( {{F_1}{\partial _\mu }\phi {\partial _\nu }\phi } \Big)\\
& - {g_{\mu \nu }}{\nabla ^\rho }{\nabla ^\sigma }\Big( {{F_1}{\partial _\rho }\phi {\partial _\sigma }\phi } \Big),
\end{aligned}
\ee
and for the variation of the GB we find the expression, valid in four dimensions
\be\label{tgb}
\begin{aligned}
T_{\mu \nu }^{GB} = &-4\Big([{\nabla _\nu }{\nabla _\mu }{F_2}]R - {g_{\mu \nu }}[{\nabla ^\sigma }{\nabla _\sigma }{F_2}]R - 2[{\nabla ^\phi }{\nabla _\mu }{F_2}]{R_{\phi \nu }} - 2[{\nabla ^\phi }{\nabla _\nu }{F_2}]{R_{\phi \mu }}\nonumber\\
&  + 2[{\nabla ^\lambda }{\nabla _\lambda }{F_2}]{R_{\mu \nu }} + 2{g_{\mu \nu }}[{\nabla ^\phi }{\nabla ^\gamma }{F_2}]{R_{\phi \gamma }} - 2[{\nabla ^\sigma }{\nabla ^\phi }{F_2}]{R_{\mu \phi \nu \sigma }}\Big).
\end{aligned}
\ee
Taking into account the variations of all the terms in the action (\ref{eq1c}) we can write the generalized Einstein equations in an arbitrary background as
\be\label{geeq}
\begin{aligned}
F(\phi)G_{\mu\nu}&=\partial_{\mu}\phi\partial_{\nu}\phi-\frac{1}{2}g_{\mu\nu}\partial_{\rho}\phi\partial^{\rho}\phi-g_{\mu\nu}V(\phi)-g_{\mu\nu}\nabla_{\sigma}\nabla^{\sigma}f(\phi)+\nabla_{\mu}\nabla_{\nu}f(\phi)\\&
{F_1}{\partial _\rho }\phi {\partial ^\rho }\phi \left( {{R_{\mu \nu }} - \frac{1}{2}{g_{\mu \nu }}R} \right) + {g_{\mu \nu }}{\nabla ^\sigma }{\nabla _\sigma }\Big( {{F_1}{\partial _\rho }\phi {\partial ^\rho }\phi } \Big) - {\nabla _\nu }{\nabla _\mu }\Big( {{F_1}{\partial _\rho }\phi {\partial ^\rho }\phi } \Big)\\
& + {F_1}R{\partial _\mu }\phi {\partial _\nu }\phi  - 2{F_1}\Big( {{R_{\mu \rho }}{\partial _\nu }\phi {\partial ^\rho }\phi  + {R_{\nu \rho }}{\partial _\mu }\phi {\partial ^\rho }\phi } \Big) + {F_1}{g_{\mu \nu }}{R_{\rho \sigma }}{\partial ^\rho }\phi {\partial ^\sigma }\phi \\
& + {\nabla ^\rho }{\nabla _\mu }\Big( {{F_1}{\partial _\nu }\phi {\partial _\rho }\phi } \Big) + {\nabla ^\rho }{\nabla _\nu }\Big( {{F_1}{\partial _\mu }\phi {\partial _\rho }\phi } \Big) - {\nabla ^\sigma }{\nabla _\sigma }\Big( {{F_1}{\partial _\mu }\phi {\partial _\nu }\phi } \Big)\\
& - {g_{\mu \nu }}{\nabla ^\rho }{\nabla ^\sigma }\Big( {{F_1}{\partial _\rho }\phi {\partial _\sigma }\phi } \Big)\\&
-4\Big([{\nabla _\nu }{\nabla _\mu }{F_2}]R - {g_{\mu \nu }}[{\nabla ^\sigma }{\nabla _\sigma }{F_2}]R - 2[{\nabla ^\phi }{\nabla _\mu }{F_2}]{R_{\phi \nu }} - 2[{\nabla ^\phi }{\nabla _\nu }{F_2}]{R_{\phi \mu }}\\
&  + 2[{\nabla ^\lambda }{\nabla _\lambda }{F_2}]{R_{\mu \nu }} + 2{g_{\mu \nu }}[{\nabla ^\phi }{\nabla ^\gamma }{F_2}]{R_{\phi \gamma }} - 2[{\nabla ^\sigma }{\nabla ^\phi }{F_2}]{R_{\mu \phi \nu \sigma }}\Big).
\end{aligned}
\ee

\section{First order perturbations of the field equations in the Newtonian gauge}

\noindent Notice that in compact notation and using the non-minimal coupling $F(\phi)$ (instead of $f(\phi)$) as it appears in the action (\ref{eq1c})
we can write the field equations, after variation of (\ref{eq1c}) with respect to the metric, as
\be\label{c1}
T^{NMC}_{\mu\nu}+T^{\phi}_{\mu\nu}+ T^{K}_{\mu\nu}+ T^{GB}_{\mu\nu}=0
\ee
where $T^{NMC}_{\mu\nu}$ is now defined as the energy momentum tensor for the action
\be\label{c2}
S_{NMC}=\int \sqrt{-g}F(\phi)R.
\ee
Expanding the equation (\ref{c1}) on the perturbed metric (\ref{eqa1}), up to first order we find
\be\label{c3}
\tilde{T}^{NMC}_{\mu\nu}+\tilde{T}^{\phi}_{\mu\nu}+ \tilde{T}^{K}_{\mu\nu}+ \tilde{T}^{GB}_{\mu\nu}+\delta T^{NMC}_{\mu\nu}+\delta T^{\phi}_{\mu\nu}+ \delta T^{K}_{\mu\nu}+ \delta T^{GB}_{\mu\nu}=0
\ee
where "tilde"corresponds to the expressions evaluated on the background metric. Then the first order perturbations of the field equations satisfy the following equation
\be\label{c3a}
\delta T_\nu ^{\mu (\phi )} + \delta T_\nu ^{\mu (NMC)} + \delta T_\nu ^{\mu (GB)}+ \delta T_\nu ^{\mu (K)}=0.
\ee
And now we use the Newtonian gauge to write the perturbations for the energy-momentum tensors.  For $\delta T_\nu ^{\mu (\phi )}$ we find
\be\label{c3b}
\begin{gathered}
  \delta T_0^{0(\phi )} = {{\dot \phi }^2}\Phi  - \dot \phi \delta \dot \phi  - V'\delta \phi  \hfill \\
  \delta T_i^{0(\phi )} = {\partial _i}\left( { - \dot \phi \delta \phi } \right) \hfill \\
  \delta T_j^{i(\phi )} - \frac{1}{3}\delta _j^i\delta T_k^{k(\phi )} = 0 \hfill \\
  \delta T_k^{k(\phi )} - \delta T_0^{0(\phi )} =  - 4\Phi{{\dot \phi }^2} + 4\dot \phi \delta \dot \phi  - 2V'\delta \phi.  \hfill \\ 
\end{gathered}
\ee

\noindent For ${\bf \delta T_\nu ^{\mu (NM)}}$
\be\label{c3c}
\delta T_0^{0(NM)} =  - 2F\left( {H(3H\Phi  + 3\dot \Psi ) -\frac{1}{a^2} \nabla^2\Psi } \right) - \dot F(3\dot \Psi  + 6H\Phi ) + 3{H^2}\delta F + 3H\delta \dot F - \frac{1}{a^2}\nabla^2\delta F,
\ee
\be\label{c3d}
\delta T_i^{0(NM)} = {\partial _i}\left( {2F(H\Phi  + \dot \Psi ) + \dot F\Phi  - \delta \dot F + H\delta F} \right),
\ee
\be\label{c3e}
\delta T_j^{i(NM)} - \frac{1}{3}\delta _j^i\delta T_k^{k(NM)} = \frac{1}{{{a^2}}}\left( {{\partial _i}{\partial _j} - \frac{1}{3}{\delta _{ij}}\nabla^2 } \right)\left( {F\left( { - \Psi  + \Phi } \right) + \delta F} \right).
\ee
\be\label{c3f}
\begin{aligned}
\delta T_k^{k(NM)} - \delta T_0^{0(NM)} =&  - 2F\left( {(3H\dot \Phi  + 3\ddot \Psi ) + 2H(3H\Phi  + 3\dot \Psi ) + 6\dot H\Phi  + \frac{1}{a^2}\nabla^2\Phi } \right)\\
 & - \dot F(3\dot \Psi  + 6H\Phi ) - 3\dot F\dot \Phi  - 6\ddot F\Phi  + 6(\dot H + {H^2})\delta F + 3\delta \ddot F \\&
 + 3H\delta \dot F -\frac{1}{a^2}\nabla^2\delta F.
\end{aligned}
\ee

\noindent For ${\bf \delta T_\nu ^{\mu (K)}}$
\be\label{c3g}
\delta T_0^{0(K)} =  - 2\dot \phi \left( { - {F_1}\dot \phi \left(  - \frac{1}{a^2}\nabla^2\Psi  + 18\Phi {H^2} + 9H\dot \Psi ) \right) - \frac{2}{a^2}{F_1}H\nabla^2\delta \phi  + 9{H^2}{F_1}\delta \dot \phi  + \frac{9}{2}{H^2}\dot \phi \delta {F_1}} \right),
\ee
\be\label{c3h}
\delta T_i^{0(K)} = {\partial _i}\left[ { - 2\dot \phi \left( { - 2H{F_1}\delta \dot \phi  + 3{H^2}{F_1}\delta \phi  - H\dot \phi \delta {F_1} + {F_1}\dot \phi \left( {\dot \Psi  + 3H\Phi } \right)} \right)} \right],
\ee
\be\label{c3i}
\delta T_j^{i(K)} - \frac{1}{3}\delta _j^i\delta T_k^{k(K)} = \frac{1}{{{a^2}}}\left( {{\partial _i}{\partial _j} - \frac{1}{3}{\delta _{ij}}\nabla^2 } \right)[ - {{\dot \phi }^2}\delta {F_1} - 2({F_1}\ddot \phi  + H{F_1}\dot \phi )\delta \phi  + {F_1}{{\dot \phi }^2}( - \Psi  - \Phi )],
\ee
\be\label{c3j}
\begin{aligned}
\delta T_k^{k(K)} - \delta T_0^{0(K)} =&  - 12H{{\dot F}_1}\dot \phi \delta \dot \phi  - 12\dot H{F_1}\dot \phi \delta \dot \phi  - 12H{F_1}\dot \phi \delta \ddot \phi  + \frac{2}{a^2}{F_1}{{\dot \phi }^2}\nabla^2\Phi  + \frac{2}{a^2}{{\dot \phi }^2}\nabla^2\delta {F_1} \\
&  + \frac{4}{a^2}{F_1}{{\dot \phi }^2}\nabla^2\Psi 
 - 6\dot H{{\dot \phi }^2}\delta {F_1} - 6H{{\dot \phi }^2}\delta {{\dot F}_1} + 2{{\dot F}_1}{{\dot \phi }^2}(12H\Phi  + 3\dot \Psi )  \\
&+ 2{F_1}{{\dot \phi }^2}(12\dot H\Phi  + 9H\dot \Phi  + 3\ddot \Psi )+ \frac{4}{a^2}{F_1}\ddot \phi \nabla^2\delta \phi  - 12{F_1}H\ddot \phi \delta \dot \phi \\
&  - 12H\dot \phi \ddot \phi \delta {F_1} + 4{F_1}\dot \phi \ddot \phi (12H\Phi  + 3\dot \Psi )
\end{aligned}
\ee
\noindent For ${\bf \delta T_\nu ^{\mu (GB)}}$\\

The perturbations of the GB energy momentum tensor from (\ref{tgb})  are given by
\be\label{eq11d}
\begin{aligned}
\delta T^{\mu GB}_{\nu} &=4\Big(\delta\left[\nabla^{\mu}\nabla_{\nu}f(\phi)\right] R+\left[\nabla^{\mu}\nabla_{\nu}f(\phi)\right] \delta R-\delta\left[\nabla^{\rho}\nabla_{\rho}f(\phi)\right]\delta^{\mu}_{\nu}R-\left[\nabla^{\rho}\nabla_{\rho}f(\phi)\right]\delta^{\mu}_{\nu}\delta R\\&
-2\delta\left[\nabla^{\mu}\nabla^{\rho}f(\phi)\right]R_{\nu\rho}-2\left[\nabla^{\mu}\nabla^{\rho}f(\phi)\right]\delta R_{\nu\rho}
-2\delta\left[\nabla^{\rho}\nabla_{\nu}f(\phi)\right]R^{\mu}_{\rho}-2\left[\nabla^{\rho}\nabla_{\nu}f(\phi)\right]\delta R^{\mu}_{\rho}\\&
+2\delta\left[\nabla^{\rho}\nabla_{\rho}f(\phi)\right]R^{\mu}_{\nu}+2\left[\nabla^{\rho}\nabla_{\rho}f(\phi)\right]\delta R^{\mu}_{\nu}+2\delta\left[\nabla^{\rho}\nabla^{\sigma}f(\phi)\right]\delta^{\mu}_{\nu}R_{\rho\sigma}+\\&2\left[\nabla^{\rho}\nabla^{\sigma}f(\phi)\right]\delta^{\mu}_{\nu}\delta R_{\rho\sigma}
-2\delta\left[\nabla^{\rho}\nabla^{\sigma}f(\phi)\right]R^{\mu}_{\rho\nu\sigma}-2\left[\nabla^{\rho}\nabla^{\sigma}f(\phi)\right]\delta R^{\mu}_{\rho\nu\sigma}\Big).
\end{aligned}
\ee
Then using (\ref{eqb21}) and the components (\ref{eqb22})-(\ref{eqb31}) we find after the corresponding simplifications 
\be\label{eq12d}
\delta T^{0 GB}_0= 24H^3\dot{\delta f(\phi)}-96H^3\dot{f(\phi)}\Phi-72H^2\dot{f(\phi)}\dot{\Psi}-\frac{8}{a^2}H^2\nabla^2\delta f(\phi)+\frac{16}{a^2}H\dot{f(\phi)}\nabla^2\Psi,
\ee
\be\label{eq13d}
\begin{aligned}
\delta T^{i GB}_j &=\frac{8}{a^2}\partial_i\partial_j\left[-\left(f''(\phi)\dot{\phi}^2+f'(\phi)\ddot{\phi}\right)\Psi+Hf'(\phi)\dot{\phi}\Phi+\left(H^2+\dot{H}\right)f'(\phi)\delta\phi\right]\\&
+8\Big[H^2\ddot{\delta f(\phi)}-\frac{1}{a^2}H^2f'(\phi)\nabla^2\delta\phi-\frac{1}{a^2}\dot{H}f'(\phi)\nabla^2\delta\phi+2H^3\dot{\delta f(\phi)}+2H\dot{H}\dot{\delta f(\phi)} \\& -\frac{1}{a^2}H\dot{f(\phi)}\nabla^2\Phi-
8H^3\dot{f(\phi)}\Phi-8H\dot{H}\dot{f(\phi)}\Phi+\frac{1}{a^2}\ddot{f(\phi)}\nabla^2\Psi-4H^2\ddot{f(\phi)}\Phi \\& -3H^2\dot{f(\phi)}\dot{\Phi}-6H^2\dot{f(\phi)}\dot{\Psi}
-2\dot{H}\dot{f(\phi)}\dot{\Psi}-2H\ddot{f(\phi)}\dot{\Psi}-2H\dot{f(\phi)}\ddot{\Psi}\Big]\delta_{ij},
\end{aligned}
\ee
\be\label{eq14d}
\begin{aligned}
\delta T^{k GB}_k &=\frac{16}{a^2}\ddot{f(\phi)}\nabla^2\Psi-\frac{16}{a^2}H\dot{f(\phi)}\nabla^2\Phi-\frac{16}{a^2}\left(H^2+\dot{H}\right)\nabla^2\delta f(\phi)+24H^2\ddot{\delta f(\phi)}+\\&
48H^3\dot{\delta f(\phi)}+48H\dot{H}\dot{\delta f(\phi)}-192H^3\dot{f(\phi)}\Phi-192H\dot{H}\dot{f(\phi)}\Phi-96H^2\ddot{f(\phi)}\Phi 
\\&-72H^2\dot{f(\phi)}\dot{\Phi}
-144H^2\dot{f(\phi)}\dot{\Psi}-48\dot{H}\dot{f(\phi)}\dot{\Psi}-48H\ddot{f(\phi)}\dot{\Psi}-48H\dot{f(\phi)}\ddot{\Psi},
\end{aligned}
\ee
\be\label{eq15d}
\delta T^{0 GB}_i =8\partial_i\left[H^3\delta f(\phi)-H^2\dot{\delta f(\phi)}+2H\dot{f(\phi)}\dot{\Psi}+3H^2\dot{f(\phi)}\Phi \right],
\ee
\be\label{eq16d}
\delta T^{i GB}_0 =\frac{8}{a^2}\partial_i\left[H^2\dot{\delta f(\phi)}-H^3\delta f(\phi)-2H\dot{f(\phi)}\dot{\Psi}-3H^2\dot{f(\phi)}\Phi \right].
\ee
\section{First order perturbations for the scalar field equation of  motion.}

\noindent From the action (\ref{eq1c}) we find the equation of motion for the scalar field as
\begin{equation}\label{eqmov}
\frac{1}{2}F'(\phi )R + {\nabla _\mu }{\nabla ^\mu }\phi  - V'(\phi ) - {F_1}'(\phi ){G_{\mu \nu }}{\nabla ^\mu }\phi {\nabla ^\nu }\phi  - 2{F_1}(\phi ){G_{\mu \nu }}{\nabla ^\mu }{\nabla ^\nu }\phi  - {F_2}'(\phi ){\cal G} = 0
\end{equation}

\noindent In order to calculate the perturbation of this equation we need the perturbation of the GB invariant, which can be evaluated as follows
\be\label{eq7dd}
\begin{aligned}
\delta{\cal G}=&2R\delta R-8\delta g^{\mu\rho}g^{\nu\sigma}R_{\mu\nu}R_{\rho\sigma}-8g^{\mu\rho}g^{\nu\sigma}\delta R_{\mu\nu}R_{\rho\sigma}-4\delta g^{\mu\alpha}g^{\sigma\delta}R^{\nu}_{\alpha\gamma\delta}R^{\gamma}_{\sigma\mu\nu}\\&
+2g^{\nu\beta}g^{\rho\gamma}g^{\sigma\delta}\delta g_{\alpha\eta}R^{\eta}_{\beta\gamma\delta}R^{\alpha}_{\nu\rho\sigma}-2 g^{\rho\gamma}g^{\sigma\delta}\delta R^{\mu}_{\beta\gamma\delta}R^{\beta}_{\mu\rho\sigma}.
\end{aligned}
\ee
Using the expressions for the perturbation of the metric (\ref{eqb2}) and of the curvatures (\ref{eqb7})-(\ref{eqb11}) and (\ref{eqb12})-(\ref{eqb18}) in the Newtonian gauge, and after some algebra we find
\be\label{eq8dd}
\begin{aligned}
\delta{\cal G}=&-\frac{8}{a^2}H^2\nabla^2\Phi+\frac{16}{a^2}H^2\nabla^2\Psi+\frac{16}{a^2}\dot{H}\nabla^2\Psi-96H^4\Phi-96H^2\dot{H}\Phi-24H^3\dot{\Phi}\\& -96H^3\dot{\Psi}-48H\dot{H}\dot{\Psi}-24H^2\ddot{\Psi}
\end{aligned}
\ee
The perturbations of the Einstein tensor, using (\ref{eqb15})-(\ref{eqb18}), are given by
\be\label{eq8d}
\delta G^0_0=6H\dot{\Psi}+6H^2\Phi-\frac{2}{a^2}\nabla^2\Psi
\ee
\be\label{eq9d}
\delta G^0_i=-2\partial_i\left(\dot{\Psi}+H\Phi\right)
\ee
\be\label{eq10d}
\delta G^i_j=\left(2\ddot{\Psi}+4\dot{H}\Phi+2H\dot{\Phi}+6H\dot{\Psi}+6H^2\Phi+\frac{1}{a^2}\nabla^2\left(\Phi-\Psi\right)\right)-\frac{1}{a^2}\partial_i\partial_j\left(\Phi-\Psi\right)
\ee

\noindent Using the above results, the first-order perturbation for the equation of motion of the scalar field (\ref{eqmov}), in the Newtonian gauge takes the form

\be\label{peqmv}
\begin{aligned}
&3H\delta \dot \phi  + 18{F_1}{H^3}\delta \dot \phi  + 12{F_1}H\dot H\delta \dot \phi  + \delta \ddot \phi  + 6{F_1}{H^2}\delta \ddot \phi  - 6H\dot \phi \Phi  - 72{F_1}{H^3}\dot \phi \Phi \\
& - 48{F_1}H\dot H\dot \phi \Phi - 2\ddot \phi \Phi  - 24{F_1}{H^2}\ddot \phi \Phi  - \dot \phi \dot \Phi  - 18{F_1}{H^2}\dot \phi \dot \Phi  - 3\dot \phi \dot \Psi  - 54{F_1}{H^2}\dot \phi \dot \Psi \\
&  - 12{F_1}\dot H\dot \phi \dot \Psi - 12{F_1}H\ddot \phi \dot \Psi  - 12{F_1}H\dot \phi \ddot \Psi  - \frac{1}{a^2}\nabla^2 \delta \phi 
- \frac{6}{a^2}{F_1}{H^2}\nabla^2 \delta \phi - \frac{4}{a^2}{F_1}\dot H\nabla^2 \delta \phi \\
&- \frac{4}{a^2}{F_1}H\dot \phi \nabla^2 \Phi  + \frac{4}{a^2}{F_1}H\dot \phi \nabla^2 \Psi + \frac{4}{a^2}{F_1}\ddot \phi \nabla^2 \Psi + 12{H^2}\Phi F' + 6\dot H\Phi F' + 3H\dot \Phi F' \\
&+ 12H\dot \Psi F' + 3\ddot \Psi F' +\frac{1}{a^2}\nabla^2 \Phi F' - \frac{2}{a^2}\nabla^2 \Psi F' + 18{H^3}\dot \phi \delta \phi {F_1}' + 12H\dot H\dot \phi \delta \phi {F_1}' \\
&+ 6{H^2}\ddot \phi \delta \phi {F_1}' + 6{H^2}\dot \phi \delta \dot \phi {F_1}' - 12{H^2}{{\dot \phi }^2}\Phi {F_1}' - 6H{{\dot \phi }^2}\dot \Psi {F_1}' + \frac{2}{a^2}\dot \phi^2\nabla^2 \Psi {F_1}' - 96{H^4}\Phi {F_2}' \\
& - 96{H^2}\dot H\Phi {F_2}' - 24{H^3}\dot \Phi {F_2}' - 96{H^3}\dot \Psi {F_2}' - 48H\dot H\dot \Psi {F_2}' - 24{H^2}\ddot \Psi {F_2}'- \frac{8}{a^2}{H^2}\nabla^2 \Phi {F_2}' \\
&+ \frac{16}{a^2}{H^2}\nabla^2 \Psi {F_2}' + \frac{16}{a^2}\dot H\nabla^2 \Psi {F_2}' - 6{H^2}\delta \phi F'' - 3\dot H\delta \phi F'' + 3{H^2}{{\dot \phi }^2}\delta \phi {F_1}'' \\
&+ 24{H^4}\delta \phi {F_2}'' +24{H^2}\dot H\delta \phi {F_2}'' + \delta \phi V'' = 0
\end{aligned}
\ee
\section{Second order action for the cosmological perturbations.}
In this section we briefly show the use of the tool {\it Xpand} \cite{cyril, martin, cyril1} to verify the results of the second-order action as presented in \cite{kobayashi1} and apply this tool to find the second order action for the model (\ref{eqm1})
. We use the gauge of the uniform field and the expression for the perturbed metric
\[d{s^2} =  - {N^2}d{t^2} + {\gamma _{ij}}(d{x^i} + {N^i}dt)(d{x^j} + {N^j}dt)\]
where
\[N = 1 + A ,\,\,\,\,\,\,\,\,{N^i} = {\partial ^i}B ,\,\,\,\,\,\,\,\,\,\,\,{\gamma _{ij}} = {a^2}(t){e^{2\xi }}\left( {{\delta _{ij}} + {h_{ij}} + \frac{1}{2}{h_{ik}}{h_{kj}}} \right)\]
with  $A$, $B$ and $\xi$ scalar perturbations and ${h_{ij}}$ the tensor perturbation satisfying ${h_{ii}}=0$, ${h_{ij}}={h_{ji}}$ y  ${{\partial _i}}{h_{ij}}=0$. Let us first focus in the scalar case (${h_{ij}}=0$). The above metric can be implemented in  {\it Xpand} \cite{cyril} as follows:
{\small
{\tt <<\text{xAct$\grave{ }$xPand$\grave{ }$};} \\
\indent{\tt DefManifold[ M, 4, \{$\alpha$, $\beta$, $\gamma$, $\mu$, $\nu$, $\lambda$, $\sigma$\} ];} \\
\indent{\tt DefMetric[ -1, g[-$\alpha$,-$\beta$], CD, PrintAs $\to$ "g" ];} \\
\indent{\tt SetSlicing[ g, n, h, cd, \{"|", "D"\}, "FLFlat" ];} \\
\indent{\tt DefMetricFields[ g, dg, h ];} \\
\indent{\tt DefMatterFields[u, du, h ];} \\
\indent{\tt \$ConformalTime = False;}\\
\indent{\tt MyMetricRules = \{dg[LI[1],$-\mu$\_,$-\nu$\_] :> - 2n[$-\mu$]n[$-\nu$]$\phi$h[LI[1]]}\\
\indent{\tt -ah[](n[$-\nu$]cd[$-\mu$]@Bsh[LI[1]]+n[$-\mu$]cd[$-\nu$]@Bsh[LI[1]])}\\
\indent{\tt +2h[$-\mu$,$-\nu$]$\psi$h[LI[1]], dg[LI[2],$-\mu$\_,$-\nu$\_] :> - 2n[$-\mu$]n[$-\nu$]$\phi$h[LI[1]]\^{}2}\\
\indent{\tt +2ah[]\^{}2 Module[\{$\alpha$\}, n[$-\mu$]n[$-\nu$]cd[$-\alpha$]@Bsh[LI[1]]cd[$\alpha$]@Bsh[LI[1]]] }\\
\indent{\tt -4ah[]$\phi$h[LI[1]](n[$-\nu$]cd[$-\mu$]@Bsh[LI[1]]+n[$-\mu$]cd[$-\nu$]@Bsh[LI[1]])}\\
\indent{\tt +4h[$-\mu$,$-\nu$]$\psi$h[LI[1]]\^{}2\};}\\
\indent{\tt kill1[expr\_]:=expr /.xAct$\grave{ }$xPand$\grave{ }$$\varphi$[xAct$\grave{ }$xTensor$\grave{ }$LI[1], xAct$\grave{ }$xTensor$\grave{ }$LI[\_]]:>0;}\\
\indent{\tt kill2[expr\_]:=expr /.xAct$\grave{ }$xPand$\grave{ }$$\varphi$[xAct$\grave{ }$xTensor$\grave{ }$LI[2], xAct$\grave{ }$xTensor$\grave{ }$LI[\_]]:>0;}\\
}
where the scalar perturbations $A$, $B$ and $\xi$ are implemented with {\tt $\phi$h}, {\tt Bsh} y {\tt $\psi$h}, respectively, and the scalar field $\phi$ is implemented with $\varphi$. The set of rules {\tt MyMetricRules} allows the reconstruction of metric perturbations  and the functions {\tt kill1} y  {\tt kill2} cancel the scalar field fluctuations (uniform field gauge). The Lagrangian density is found in the following way:

{\small
{\tt DefScalarFunction[V];} \\
\indent{\tt DefScalarFunction[F];}\\
\indent{\tt DefScalarFunction[F1];}\\
\indent{\tt DefScalarFunction[F2];}\\
\indent{\tt Lag = kill2[kill1[ExpandPerturbation@Perturbed[Conformal[g, gah2]}\\
\indent{\tt \big[Sqrt[-Detg[]]\big((1/2)F[$\varphi$[]]RicciScalarCD[] - (1/2)CD[$-\mu$][$\varphi$[]]CD[$\mu$][$\varphi$[]]    }\\
\indent{\tt - V[$\varphi$[]] + F1[$\varphi$[]]EinsteinCD[$-\mu$,$-\nu$]CD[$\mu$][$\varphi$[]]CD[$\nu$][$\varphi$[]] - F2[$\varphi$[]](  }\\
\indent{\tt RicciScalarCD[]\^{}2 - 4 RicciCD[-$\alpha$,-$\beta$] RicciCD[$\alpha$,$\beta$]    }\\
\indent{\tt + 
  RiemannCD[-$\alpha$,-$\beta$, -$\gamma$,-$\lambda$] RiemannCD[$\alpha$,$\beta$, $\gamma$,$\lambda$])\big)\big], 2]]];}\\
\indent{\tt ExtractOrder[ExtractComponents[SplitPerturbations[Lag, MyMetricRules, h]}\\
\indent{\tt , h], 2]//Expand;}
}\\
canceling a large number of the boundary terms and using Eqs. (\ref{eqm5}) y (\ref{eqm6}), the result obtained with {\it Xpand} can be reduced to:
\begin{align}\label{kob1}
\delta S_s^2 = \int dt{d^3}x{a^3}\Big[& - 3{G_T} {{\dot \xi }^2} + \frac{{{F_T}}}{{{a^2}}}{\partial _i}\xi {\partial _i}\xi  + \Sigma {A^2} - 2\Theta A\frac{{{\partial _i}{\partial _i}B}}{{{a^2}}} + 2{G_T}\dot \xi \frac{{{\partial _i}{\partial _i}B}}{{{a^2}}}\nonumber\\
&+6\Theta A\dot \xi  - 2{G_T}A\frac{{{\partial _i} {\partial _i}\xi }}{{{a^2}}}\Big]
\end{align}
where 
\be\label{slr4}
{ G}_T=F-F_1\dot{\phi}^2-8H\dot{F}_2.
\ee
\be\label{slr5}
{ F}_T=F+F_1\dot{\phi}^2-8\ddot{F}_2
\ee
\be\label{slr6}
\Theta=FH+\frac{1}{2}\dot{F}-3HF_1\dot{\phi}^2-12H^2\dot{F}_2
\ee
\be\label{slr7}
\Sigma=-3FH^2-3H\dot{F}+\frac{1}{2}\dot{\phi}^2+18H^2F_1\dot{\phi}^2+48H^3\dot {F}_2
\ee
From (\ref{kob1})  it is easy to obtain the equations of motion for $A$ and $B$, which are given by
\begin{equation}\label{restric1}
\Sigma A + 3\Theta \dot \xi  - \Theta \frac{{{\partial _i}{\partial _i}B}}{{{a^2}}} - {G_T}\frac{{{\partial _i}{\partial _i}\xi }}{{{a^2}}} = 0
\end{equation}
\begin{equation}\label{restric2}
A = \frac{{{G_T}}}{\Theta }\dot \xi 
\end{equation}
By replacing the equation (\ref{restric2}) in (\ref{restric1}) it is obtained
\begin{equation}\label{restric3}
\frac{{{\partial _i}{\partial _i}B}}{{{a^2}}} = \frac{\Sigma }{{{\Theta ^2}}}{G_T}\dot \xi  + 3\dot \xi  - \frac{{{G_T}}}{\Theta }\frac{{{\partial _i}{\partial _i}\xi }}{{{a^2}}}
\end{equation}
Replacing Eqs. (\ref{restric2}) and (\ref{restric3})  in (\ref{kob1}) after simplifying it is obtained:
\[\delta S_s^2 = \int {dt{d^3}x{a^3}\left[ {\left( {3{G_T} + \Sigma {{\left( {\frac{{{G_T}}}{\Theta }} \right)}^2}} \right){{\dot \xi }^2} + \frac{{{F_T}}}{{{a^2}}}{\partial _i}\xi {\partial _i}\xi  - 2\frac{{G_T^2}}{\Theta }\dot \xi \frac{{{\partial _i}{\partial _i}\xi }}{{{a^2}}}} \right]} \]
Omitting total spatial derivatives in the last term, the previous expression can be rewritten as 
\begin{equation}\label{kob2}
\delta S_s^2 = \int {dt{d^3}x{a^3}\left[ {\left( {3{G_T} + \Sigma {{\left( {\frac{{{G_T}}}{\Theta }} \right)}^2}} \right){{\dot \xi }^2} + \frac{{{F_T}}}{{{a^2}}}{\partial _i}\xi {\partial _i}\xi  + 2\frac{{G_T^2}}{\Theta }\frac{{{\partial _i}\dot \xi {\partial _i}\xi }}{{{a^2}}}} \right]} 
\end{equation}
From
\[\frac{d}{{dt}}\left[ {a\frac{{G_T^2}}{\Theta }{\partial _i}\xi {\partial _i}\xi } \right] = \frac{d}{{dt}}\left[ {a\frac{{G_T^2}}{\Theta }} \right]{\partial _i}\xi {\partial _i}\xi  + 2a\frac{{G_T^2}}{\Theta }{\partial _i}\dot \xi {\partial _i}\xi, \]
it follows that the last tern of (\ref{kob2}) can be rewritten by using the previous expression (omitting total derivative). In this manner one obtains:
\[\delta S_s^2 = \int {dt{d^3}x\left[ {{a^3}\left( {\left( {3{G_T} + \Sigma {{\left( {\frac{{{G_T}}}{\Theta }} \right)}^2}} \right){{\dot \xi }^2} + \frac{{{F_T}}}{{{a^2}}}{\partial _i}\xi {\partial _i}\xi } \right) - \frac{d}{{dt}}\left( {a\frac{{G_T^2}}{\Theta }} \right){\partial _i}\xi {\partial _i}\xi } \right]} \]
Organizing terms this expression takes the form 
\[\delta S_s^2 = \int {dt{d^3}x{a^3}\left[ {\left( {3{G_T} + \Sigma {{\left( {\frac{{{G_T}}}{\Theta }} \right)}^2}} \right){{\dot \xi }^2} - \frac{1}{{{a^2}}}\left( {\frac{1}{a}\frac{d}{{dt}}\left( {a\frac{{G_T^2}}{\Theta }} \right) - {F_T}} \right){\partial _i}\xi {\partial _i}\xi } \right]} \]
Defining the quantities 
\[\begin{gathered}
  {F_s} = \frac{1}{a}\frac{d}{{dt}}\left( {a\frac{{G_T^2}}{\Theta }} \right) - {F_T} \hfill \\
  {G_s} = 3{G_T} + \Sigma {\left( {\frac{{{G_T}}}{\Theta }} \right)^2}, \hfill \\ 
\end{gathered} \]
The expression for the second order action takes the form
\[\delta S_s^2 = \int {dt{d^3}x{a^3}\left[ {{G_s}{{\dot \xi }^2} - \frac{{{F_s}}}{{{a^2}}}{\partial _i}\xi {\partial _i}\xi } \right]}  = \int {dt{d^3}x{a^3}{G_s}\left[ {{{\dot \xi }^2} - \frac{{c_s^2}}{{{a^2}}}{\partial _i}\xi {\partial _i}\xi } \right]} \]
where 
\[c_s^2 = \frac{{{F_s}}}{{{G_s}}}\]

\noindent In order to implement the {\bf tensor perturbations}  $h_{ij}$ in the {\it Xpand} algorithm, the function {\tt MyMetricRules} must be modified as follows:

{\small
\indent{\tt MyMetricRules = \{dg[LI[1],$-\mu$\_,$-\nu$\_] :> Eth[LI[1],$-\mu$,$-\nu$],}\\
\indent{\tt dg[LI[2],$-\mu$\_,$-\nu$\_] :> Module[\{$\alpha$\}, Eth[LI[1],$-\mu$,$\alpha$]Eth[LI[1],$-\nu$,$-\alpha$]]\};}\\}
where the fluctuations $h_{ij}$ are implemented with {\tt Eth}. The explicit expression obtained from the algorithm for the second-order action is:
\begin{align}
\delta S_T^2 =& \int {dt{d^3}x\Bigg[\frac{{{a^3}{{\dot h}_{ij}}{{\dot h}_{ij}}F}}{8} + 2{a^3}{{\dot h}_{ij}}{{\ddot h}_{ij}}{F_2}H}  + 3{a^3}{{\dot h}_{ij}}{{\dot h}_{ij}}{F_2}{H^2} + {a^3}{{\dot h}_{ij}}{{\dot h}_{ij}}{F_2}\dot H \nonumber\\
&- \frac{{{a^3}{{\dot h}_{ij}}{{\dot h}_{ij}}{F_1}{{\dot \phi }^2}}}{8} - 2a{{\ddot h}_{ij}}{F_2}{\partial _k}{\partial _k}{h_{ij}} - 4a{{\dot h}_{ij}}{F_2}H{\partial _k}{\partial _k}{h_{ij}} - \frac{{a{\partial _k}{h_{ij}}{\partial _k}{h_{ij}}F}}{8} \nonumber\\
&+ a{F_2}{H^2}{\partial _k}{h_{ij}}{\partial _k}{h_{ij}} + a{F_2}\dot H{\partial _k}{h_{ij}}{\partial _k}{h_{ij}} - \frac{{a{\partial _k}{h_{ij}}{\partial _k}{h_{ij}}{F_1}{{\dot \phi }^2}}}{8} - 2a{F_2}{\partial _j}{{\dot h}_{ik}}{\partial _k}{{\dot h}_{ij}} \nonumber\\
&+ 2a{F_2}{\partial _k}{{\dot h}_{ij}}{\partial _k}{{\dot h}_{ij}} + \frac{{{F_2}{\partial _k}{\partial _k}{h_{ij}}{\partial _l}{\partial _l}{h_{ij}}}}{a} - \frac{{{F_2}{\partial _j}{\partial _i}{h_{kl}}{\partial _l}{\partial _k}{h_{ij}}}}{a} + \frac{{2{F_2}{\partial _l}{\partial _j}{h_{ik}}{\partial _l}{\partial _k}{h_{ij}}}}{a}\nonumber\\
& - \frac{{{F_2}{\partial _l}{\partial _k}{h_{ij}}{\partial _l}{\partial _k}{h_{ij}}}}{a}\Bigg]\nonumber
\end{align}
Notice that the terms $12^\circ $, $15^\circ $ and $16^\circ $ are zero since ${{\partial _i}}{h_{ij}}=0$ (omitting surface terms). In addition, the terms $14^\circ $ y $17^\circ $ cancel each other (omitting surface terms). In this manner it is obtained
\begin{align}\label{s2tenx}
\delta S_T^2 =&\int {dt{d^3}x\Bigg[\frac{{{a^3}{{\dot h}_{ij}}{{\dot h}_{ij}}F}}{8} + 2{a^3}{{\dot h}_{ij}}{{\ddot h}_{ij}}{F_2}H}  + 3{a^3}{{\dot h}_{ij}}{{\dot h}_{ij}}{F_2}{H^2} + {a^3}{{\dot h}_{ij}}{{\dot h}_{ij}}{F_2}\dot H \nonumber\\
&- \frac{{{a^3}{{\dot h}_{ij}}{{\dot h}_{ij}}{F_1}{{\dot \phi }^2}}}{8} - 2a{{\ddot h}_{ij}}{F_2}{\partial _k}{\partial _k}{h_{ij}} - 4a{{\dot h}_{ij}}{F_2}H{\partial _k}{\partial _k}{h_{ij}} - \frac{{a{\partial _k}{h_{ij}}{\partial _k}{h_{ij}}F}}{8} \nonumber\\
& + a{F_2}{H^2}{\partial _k}{h_{ij}}{\partial _k}{h_{ij}}+ a{F_2}\dot H{\partial _k}{h_{ij}}{\partial _k}{h_{ij}} - \frac{{a{\partial _k}{h_{ij}}{\partial _k}{h_{ij}}{F_1}{{\dot \phi }^2}}}{8} + 2a{F_2}{\partial _k}{{\dot h}_{ij}}{\partial _k}{{\dot h}_{ij}}\Bigg]
\end{align}
Since
\[\frac{d}{{dt}}\left( {{a^3}{{\dot h}_{ij}}{{\dot h}_{ij}}{F_2}H} \right) = 3{a^3}{H^2}{{\dot h}_{ij}}{{\dot h}_{ij}}{F_2} + 2{a^3}{{\dot h}_{ij}}{{\ddot h}_{ij}}{F_2}H + {a^3}{{\dot h}_{ij}}{{\dot h}_{ij}}{{\dot F}_2}H + {a^3}{{\dot h}_{ij}}{{\dot h}_{ij}}{F_2}\dot H\]
then, the fourth term of (\ref{s2tenx}) can be rewritten by using the previous expression (up to total derivatives). In this way it is found:
\begin{align}\label{s2tenx1}
\delta S_T^2 =&\int {dt{d^3}x\Bigg[{a^3}\left( {\frac{F}{8} - {{\dot F}_2}H - \frac{{{F_1}{{\dot \phi }^2}}}{8}} \right)} {{\dot h}_{ij}}{{\dot h}_{ij}} - 2a{{\ddot h}_{ij}}{F_2}{\partial _k}{\partial _k}{h_{ij}} - 4a{{\dot h}_{ij}}{F_2}H{\partial _k}{\partial _k}{h_{ij}}\nonumber\\
& - \frac{{a{\partial _k}{h_{ij}}{\partial _k}{h_{ij}}F}}{8} + a{F_2}{H^2}{\partial _k}{h_{ij}}{\partial _k}{h_{ij}} + a{F_2}\dot H{\partial _k}{h_{ij}}{\partial _k}{h_{ij}} - \frac{{a{\partial _k}{h_{ij}}{\partial _k}{h_{ij}}{F_1}{{\dot \phi }^2}}}{8}\nonumber\\
& + 2a{F_2}{\partial _k}{{\dot h}_{ij}}{\partial _k}{{\dot h}_{ij}}\Bigg]
\end{align}
The third and sixth terms can be expressed, taking into account the following expressions
\begin{equation}\label{ayu1}
{\partial _k}\left( {4a{{\dot h}_{ij}}{F_2}H{\partial _k}{h_{ij}}} \right) = 4a{F_2}H{\partial _k}{{\dot h}_{ij}}{\partial _k}{h_{ij}} + 4a{{\dot h}_{ij}}{F_2}H{\partial _k}{\partial _k}{h_{ij}}
\end{equation}
\begin{align}\label{ayu1}
\frac{d}{{dt}}\left( {a{F_2}H{\partial _k}{h_{ij}}{\partial _k}{h_{ij}}} \right) =& a{F_2}{H^2}{\partial _k}{h_{ij}}{\partial _k}{h_{ij}} + a{{\dot F}_2}H{\partial _k}{h_{ij}}{\partial _k}{h_{ij}} + a{F_2}\dot H{\partial _k}{h_{ij}}{\partial _k}{h_{ij}}\nonumber\\
& + 2a{F_2}H{\partial _k}{{\dot h}_{ij}}{\partial _k}{h_{ij}}
\end{align}
Omitting surface terms and total derivatives, the Eq. (\ref{s2tenx1}) takes the form
\begin{align}\label{s2tenx2}
\delta S_T^2 =&\int {dt{d^3}x\Bigg[{a^3}\left( {\frac{F}{8} - {{\dot F}_2}H - \frac{{{F_1}{{\dot \phi }^2}}}{8}} \right)} {{\dot h}_{ij}}{{\dot h}_{ij}} - 2a{{\ddot h}_{ij}}{F_2}{\partial _k}{\partial _k}{h_{ij}} + 2a{F_2}H{\partial _k}{{\dot h}_{ij}}{\partial _k}{h_{ij}}\nonumber\\
&- \frac{{a{\partial _k}{h_{ij}}{\partial _k}{h_{ij}}F}}{8} - a{{\dot F}_2}H{\partial _k}{h_{ij}}{\partial _k}{h_{ij}} - \frac{{a{\partial _k}{h_{ij}}{\partial _k}{h_{ij}}{F_1}{{\dot \phi }^2}}}{8} + 2a{F_2}{\partial _k}{{\dot h}_{ij}}{\partial _k}{{\dot h}_{ij}}\Bigg]
\end{align}
The second term can be rewritten if  taking into account the following expressions
\[{\partial _k}\left( {2a{{\ddot h}_{ij}}{F_2}{\partial _k}{h_{ij}}} \right) = 2a{\partial _k}{{\ddot h}_{ij}}{F_2}{\partial _k}{h_{ij}} + 2a{{\ddot h}_{ij}}{F_2}{\partial _k}{\partial _k}{h_{ij}}\]
\begin{align}
\frac{d}{{dt}}\left( {a{\partial _k}{{\dot h}_{ij}}{F_2}{\partial _k}{h_{ij}}} \right) =& aH{\partial _k}{{\dot h}_{ij}}{F_2}{\partial _k}{h_{ij}} + a{\partial _k}{{\ddot h}_{ij}}{F_2}{\partial _k}{h_{ij}} + a{\partial _k}{{\dot h}_{ij}}{{\dot F}_2}{\partial _k}{h_{ij}}\nonumber\\
& + a{\partial _k}{{\dot h}_{ij}}{F_2}{\partial _k}{{\dot h}_{ij}}\nonumber
\end{align}
After which, the action (\ref{s2tenx2}) becomes
\begin{align}
\delta S_T^2 =&\int {dt{d^3}x\Bigg[{a^3}\left( {\frac{F}{8} - {{\dot F}_2}H - \frac{{{F_1}{{\dot \phi }^2}}}{8}} \right)} {{\dot h}_{ij}}{{\dot h}_{ij}} - 2a{\partial _k}{{\dot h}_{ij}}{{\dot F}_2}{\partial _k}{h_{ij}} - \frac{{a{\partial _k}{h_{ij}}{\partial _k}{h_{ij}}F}}{8}\nonumber\\
& - a{{\dot F}_2}H{\partial _k}{h_{ij}}{\partial _k}{h_{ij}} - \frac{{a{\partial _k}{h_{ij}}{\partial _k}{h_{ij}}{F_1}{{\dot \phi }^2}}}{8}\Bigg]
\end{align}
The fourth term can be rewritten if taking into account that
\[\frac{d}{{dt}}\left( {a{{\dot F}_2}{\partial _k}{h_{ij}}{\partial _k}{h_{ij}}} \right) = a{{\dot F}_2}H{\partial _k}{h_{ij}}{\partial _k}{h_{ij}} + a{{\ddot F}_2}{\partial _k}{h_{ij}}{\partial _k}{h_{ij}} + 2a{{\dot F}_2}{\partial _k}{{\dot h}_{ij}}{\partial _k}{h_{ij}}\]
using this expression and simplifying, it follows that 
\[\delta S_T^2 =\frac{1}{8}\int {dt{d^3}x{a^3}\Bigg[\left( {F - 8{{\dot F}_2}H - {F_1}{{\dot \phi }^2}} \right)} {{\dot h}_{ij}}{{\dot h}_{ij}} - \frac{1}{{{a^2}}}\left( {F - 8{{\ddot F}_2} + {F_1}{{\dot \phi }^2}} \right){\partial _k}{h_{ij}}{\partial _k}{h_{ij}}\Bigg],\]
and using the definitions (\ref{slr4}) y (\ref{slr5}), the final expression for the second-order action takes the form
\begin{equation}
\delta S_T^2 =\frac{1}{8}\int {dt{d^3}x{a^3}\Bigg[{G_T}} {{\dot h}_{ij}}{{\dot h}_{ij}} - \frac{{{F_T}}}{{{a^2}}}{\partial _k}{h_{ij}}{\partial _k}{h_{ij}}\Bigg]
\end{equation}
which gives the velocity of the tensor perturbations as
\begin{equation}
c_T^2 = \frac{{{F_T}}}{{{G_T}}}
\end{equation}

\section{The slow-roll inflation for the minimally coupled scalar field}
In general for a second order action
\be\label{f5}
S^{(2)}=\int dtd^3x a^3{\cal G}_s\left[\dot{\xi}^2-\frac{c_s^2}{a^2}(\nabla \xi)^2\right]
\ee
one finds the equation of motion of the scalar perturbation as
\be\label{f6}
\frac{d}{dt}\left(a^3{\cal G}_s\right)\dot{\xi}+a^3{\cal G}_s\ddot{\xi} - a c_s^2{\cal G}_s\nabla^2\xi=0
\ee
which can be written as
\be\label{f7}
\ddot{\xi}+\frac{1}{a^3{\cal G}_s}\frac{d}{dt}\left(a^3{\cal G}_s\right)\dot{\xi}-\frac{c_s^2}{a^2}\nabla^2\xi=0
\ee
or in Fourier modes ($\nabla^2\rightarrow -k^2$)
\be\label{f8}
\ddot{\xi}_k+\frac{1}{a^3{\cal G}_s}\frac{d}{dt}\left(a^3{\cal G}_s\right)\dot{\xi}_k+\frac{c_s^2}{a^2}k^2\xi_k=0
\ee
where $k$ is the wave number $k=2\pi/\lambda$.
For small $k$ beyond the horizon, i.e. $c_sk<<aH$ one can neglect the third term and write
\be\label{f9}
a^3{\cal G}_s\ddot{\xi}_k+\frac{d}{dt}\left(a^3{\cal G}_s\right)\dot{\xi}_k=\frac{d}{dt}\left(a^3{\cal G}_s\dot{\xi}_k\right)=0
\ee
which gives
\be\label{f10}
\dot{\xi}_k=\frac{c_k}{a^3{\cal G}_s},\;\; \rightarrow \;\;\; \xi_k=d_k+c_k \int \frac{dt}{a^3{\cal G}_s}
\ee
where $c_k$ and $d_k$ are integration constants. Note that $d_k$ corresponds to the constant (observable) mode and the integral gives the decaying mode, under the assumption that during inflation ${\cal G}_s$ is slowly varying.  To canonical normalize the curvature perturbations (\ref{f5}) we make the following change of variables 
\be\label{f11}
v_k=z\xi_k,\;\;\; z=a\sqrt{2{\cal G}_s},\;\;\; dt=a d\tau
\ee
then,
$$
\dot{\xi}=\frac{d\xi}{d\tau}\frac{d\tau}{dt}=\frac{1}{a}\xi'=\frac{1}{a}(\frac{v}{z})'=\frac{1}{az}\left(v'-\frac{z'}{z}v\right)
$$
where $'$ denotes derivative w.r.t  $\tau$. The second order action (\ref{f5}) transforms as 
$$
\begin{aligned}
S^{(2)}&=\int d\tau d^3x a^4\left[\frac{{\cal G}_s}{a^2z^2}\left(v'-\frac{z'}{z}v\right)^2-\frac{c_s^2{\cal G}_s}{a^2z^2}(\nabla v)^2\right]\\ &
=\int d\tau d^3x \frac{1}{2}\left[\left(v'-\frac{z'}{z}v\right)^2-c_s^2(\nabla V)^2\right]
\end{aligned}
$$
which finally gives after integration by parts
\be\label{f12}
S^{(2)}=\int d\tau d^3x \frac{1}{2}\left[v'^2+\frac{z''}{z}v^2-c_s^2(\nabla v)^2\right]
\ee
The equation of motion for $v$ that follows from the above action is
\be\label{f13}
v''-c_s^2\nabla^2 v- \frac{z''}{z}v=0
\ee
which is the Mukhanov equation. One can also define the Fourier expansion of the field $v$ as
\be\label{eqf13a}
v(x,\tau)=\int \frac{d^3k}{(2\pi)^3}v_k(\tau)e^{i\vec{k} . \vec{x}},
\ee
leading to
\be\label{f13b}
v_k''+\left(c_S^2k^2-\frac{z''}{z}\right)v_k=0.
\ee
The difficulty in solving this equation lies in the function $z''/z$ and the velocity of the scalar perturbations $c_S$ which encode the dynamics of the model on the given inflationary background. Nevertheless, appropriate analytical solutions can be obtained in the de Sitter limit and under the slow-roll approximation.\\

\noindent {\bf  The minimally coupled scalar field}\\

For the canonical scalar field the Friedman and field equations can be reduced to
\be\label{f14} 
H^2=\frac{1}{3M_p^2}\left(\frac{1}{2}\dot{\phi}^2+V(\phi)\right)
\ee
\be\label{f15}
\dot{H}=-\frac{1}{2M_p^2}\dot{\phi}^2
\ee
To analyze the second order action in this case we set $F=1/\kappa^2=M_p^2$, $F_1=F_2=0$ in (\ref{slr1})-(\ref{slr7}), which gives
\be\label{f16}
\Sigma=-3FH^2+\frac{1}{2}\dot{\phi}^2,\;\;\; \Theta=M_p^2 H,\;\;\; {\cal F}_T=M_p^2,\;\;\; {\cal G}_T=M_p^2.
\ee
Therefore
\be\label{f17}
 {\cal G}_s=\frac{\dot{\phi}^2}{2H^2}
 \ee
 and 
\be\label{f18}
{\cal F}_s=\frac{M_p^2}{a}\frac{d}{dt}\left(\frac{a}{H}\right)-M_p^2=-M_p^2\frac{\dot{H}}{H^2},
\ee
giving $c_s^2=1$. The second order action for this simplified case takes the form 
\be\label{f19}
S^{(2)}=\int dtd^3x a^3\frac{\dot{\phi}^2}{2H^2}\left[\dot{\xi}^2-\frac{1}{a^2}(\nabla\xi)^2\right]
\ee
which in the Mukhanov variables becomes
\be\label{f120}
S^{(2)}=\int d\tau d^3x \frac{1}{2}\left[v'^2+\frac{z''}{z}v^2-(\nabla v)^2\right]
\ee
The Mukhanov equation (\ref{f13}) for the case of canonical scalar field simplifies taking into account (\ref{f17}) and 
\be\label{f20a}
 z=a\frac{\dot{\phi}}{H}.
\ee
Then,
$$
z'=\frac{dz}{d\tau}=a\frac{dz}{dt}=a\left(\dot{a}\frac{\dot{\phi}}{H}+a\frac{\ddot{\phi}}{H}-a\frac{\dot{\phi}\dot{H}}{H^2}\right).
$$
On the other hand, the standard slow roll parameters for this case are defined as
\be\label{f21}
\epsilon=-\frac{\dot{H}}{H^2}=\frac{\dot{\phi}^2}{2M_p^2H^2}
\ee
and
\be\label{f22}
\eta=\frac{\dot{\epsilon}}{H\epsilon}=\frac{2\ddot{\phi}}{\dot{\phi}H}-\frac{\dot{H}}{H^2}=2\left(\epsilon-\delta\right)
\ee
where
\be\label{f23}
\delta=-\frac{\ddot{\phi}}{\dot{\phi}H}
\ee
so, if $\epsilon, \delta <<1$ then $\eta<<1$.\\
Then $z'/z$ may be written as
\be\label{f24}
\frac{z'}{z}=aH\left(1+\frac{\ddot{\phi}}{H\dot{\phi}}-\frac{\dot{H}}{H^2}\right)=aH\left(1-\delta+\epsilon\right)
\ee
Taking the derivative with respect to $\tau$ one finds
$$
\begin{aligned}
\frac{d}{d\tau}\left(\frac{z'}{z}\right)&=\frac{z''}{z^2}-\left(\frac{z'}{z}\right)^2=a\frac{d}{dt}\left[aH\left(1-\delta+\epsilon\right)\right]\\
&=a^2H^2\left[1-\delta+\epsilon-\epsilon\left(1-\delta+\epsilon\right)+\frac{\dot{\epsilon}}{H}-\frac{\dot{\delta}}{H}\right]\\ &\simeq a^2H^2\left[1-\delta+\epsilon-\epsilon\left(1-\delta+\epsilon\right)+\frac{\dot{\epsilon}}{H}\right]
\end{aligned}
$$
where we have used $\dot{a}=aH$ and in the last equality we have neglected $\dot{\delta}$.
Then, using (\ref{f24}), up to first order in slow roll parameters one can write 
\be\label{f25}
\frac{z''}{z^2}\simeq a^2H^2\left(2+2\epsilon-3\delta\right).
\ee
Note that form the equality 
$$
\frac{d}{d\tau}(aH)^{-1}=\epsilon-1,
$$
if we consider that $\epsilon$ varies very slowly with time, i.e. is quasi constant, then one finds
\be\label{f26}
(aH)^{-1}=\left(\epsilon-1\right)\tau \;\;\; \Rightarrow \;\;\; \tau=\frac{1}{aH}\frac{1}{(\epsilon-1)}
\ee
which is the conformal time. Note that in de Sitter $\epsilon=0$ and one has 
$$
\tau_{dS}=-\frac{1}{aH}
$$
so the comovil horizon is equal to the conformal time, and then neglecting the slow roll parameters, the following approximation 
$$
\frac{z''}{z}\simeq 2a^2H^2\left(1+\epsilon-\frac{3}{2}\delta\right)\simeq 2a^2H^2=\frac{2}{\tau_{dS}^2}
$$
takes place in de Sitter. But taking into account the slow roll parameters and using (\ref{f26}) we find
\be\label{f27}
\frac{z''}{z}\simeq \frac{1}{\tau^2}\frac{2+2\epsilon-3\delta}{(1-\epsilon)^2}=\frac{1}{\tau^2}\left(\mu^2-\frac{1}{4}\right)
\ee
where
$$
\mu^2=\frac{1}{4}+\frac{2+2\epsilon-3\delta}{(1-\epsilon)^2}\simeq \frac{9}{4}+6\epsilon-3\delta.
$$
Then 
$$
\mu \simeq \frac{3}{2}+2\epsilon-\delta
$$
Using (\ref{f27}) in the Mukhanov equation (\ref{f13b}) with $c_s^2=1$ we find in the Fourier modes
\be\label{f28}
v_k''+k^2 v_k- \frac{1}{\tau^2}\left(\mu^2-\frac{1}{4}\right)v_k=0.
\ee
First note that deep inside the horizon, when the condition $k>>aH$ or $\tau\rightarrow -\infty$ is fulfilled, the mode equation becomes
\be\label{eqf29}
v_k''+k^2 v_k=0.
\ee
which allows the quantization of the mode function in complete analogy with the quantization of (massless) scalar field on Minkowski background. Then, the choice of vacuum as the minimum energy state and the positivity of the normalization condition for the fluctuations $v_k$ \cite{riotto, mukhanov, birell, mukhanov3} leads to the unique plane-wave solution
\be\label{f30}
v_k=\frac{1}{\sqrt{2k}}e^{-ik\tau}.
\ee
This solution can be used as a boundary condition (at $k>>aH$) for the general solution of Eq. (\ref{f28}). 
Assuming $\mu^2$ constant for slowly varying slow-roll parameters, the general solution to the equation (\ref{f28}) is given by
\be\label{f31}
v_k=\sqrt{-\tau}\Big[c_{1k}H_{\mu}^{(1)}(-k\tau)+c_{2k}H_{\mu}^{(2)}(-k\tau)\Big]
\ee
where $H_{\mu}^{(1)}$ and $H_{\mu}^{(2)}$ are the Hankel functions of the first and second kind respectively. 
These functions have the following asymptotic behavior 
\be\label{f32}
H_{\mu}^{(1)}(x>>1)\simeq \sqrt{\frac{2}{\pi x}}e^{i(x-\frac{\pi}{2}\mu-\frac{\pi}{4})}
\ee
\be\label{f33}
H_{\mu}^{(2)}(x>>1)\simeq \sqrt{\frac{2}{\pi x}}e^{-i(x-\frac{\pi}{2}\mu-\frac{\pi}{4})}
\ee
Taking $x=-k\tau$, if $x>>1$ then $k>>aH$ which corresponds to sub horizon scales. 
Then imposing (\ref{f30}) as the boundary condition at $-k\tau>>1$, it is found that 
$$
c_{1k}=\frac{\sqrt{\pi}}{2}e^{i\frac{\pi}{2}(\mu+\frac{1}{2})},\;\;\; c_{2k}=0
$$
and the general solution takes the form
\be\label{f34}
v_k=\frac{\sqrt{\pi}}{2}e^{i\frac{\pi}{2}(\mu+\frac{1}{2})}\sqrt{-\tau}H_{\mu}^{(1)}(-k\tau)
\ee
On the other hand, on super horizon scales where $k<<aH$ ($x<<1$), the Hankel function has the following asymptotic behavior
\be\label{f35}
H_{\mu}^{(1)}(x)=\sqrt{\frac{2}{\pi}}e^{-i\frac{\pi}{2}}2^{\mu-\frac{3}{2}}\frac{\Gamma(\mu)}{\Gamma(\frac{3}{2})}x^{-\mu}
\ee
and replacing in (\ref{f34}) we find the solution
\be\label{f36}
v_k=e^{i\frac{\pi}{2}(\mu-\frac{1}{2})}2^{\mu-\frac{3}{2}}\frac{\Gamma(\mu)}{\Gamma(\frac{3}{2})}\frac{1}{\sqrt{2}}\sqrt{-\tau}\left(-k\tau\right)^{-\mu}
\ee
To evaluate the power spectra we find from (\ref{f24}) and (\ref{f26})
\be\label{f37}
\frac{z'}{z}=\frac{\left(1-\delta+\epsilon\right)}{\epsilon-1}\frac{1}{\tau}
\ee
for slowly varying slow roll parameters one finds
\be\label{f38}
z\propto \tau^{\frac{1}{2}-\mu}
\ee
where
$$\mu=\frac{3}{2}+2\epsilon-\delta.$$
Assuming $\mu\simeq 3/2$ in $2^{\mu-\frac{3}{2}}$ and $\Gamma(\mu)$ in (\ref{f36}) gives
\be\label{f39}
v_k=\frac{1}{\sqrt{2}}e^{i\pi/2}\sqrt{-\tau}\left(-k\tau\right)^{-\mu}.
\ee
Then in the super horizon regime 
\be\label{f40}
\xi_k=\frac{v_k}{z}\propto \tau^0 k^{-\mu}=k^{-\frac{3}{2}-2\epsilon+\delta}
\ee
depending only on $k$, which agrees with the solution $\xi_k=const.$ on super horizon scales (see (\ref{f10})). Then for the power spectra we find
\be\label{f41}
P_{\xi}=\frac{k^3}{(2\pi)^2}|\xi_k|^2\propto k^{2\delta-4\epsilon}
\ee
and the scalar spectral index is given by
\be\label{f42}
n_s-1=\frac{d \ln P(\xi)}{d \ln k}=2\delta-4\epsilon
\ee
where the scale invariance corresponds to $n_s=1$.

\section*{Acknowledgments}
\noindent This work was supported by Universidad del Valle under project CI 71074 and by
COLCIENCIAS Grant No. 110671250405. DFJ acknowledges support from COLCIENCIAS, Colombia.

\end{document}